\documentclass[12pt]{article}
\newif\ifarxiv\arxivtrue
\usepackage[utf8]{inputenc}
\usepackage[T1]{fontenc}
\usepackage[margin=1in]{geometry}
\usepackage{setspace}
\usepackage{amsmath,amssymb}
\usepackage{booktabs}
\usepackage{threeparttable}
\usepackage{graphicx}
\usepackage{array}
\usepackage{caption}
\usepackage{multirow}
\usepackage{etoolbox}
\usepackage[authoryear,round]{natbib}
\usepackage{xurl}
\usepackage[colorlinks=true, citecolor=blue, linkcolor=blue, urlcolor=blue]{hyperref}

\usepackage[htt]{hyphenat}   
\graphicspath{{./}}
\newcommand{\cw}[1]{\texttt{\small #1}}
\newcommand{\pq}[1]{``#1''}

\AtBeginEnvironment{tabular}{\setstretch{1}\small}
\AtBeginEnvironment{threeparttable}{\setstretch{1}\let\scriptsize\footnotesize}
\AtBeginEnvironment{tablenotes}{\setstretch{1}\footnotesize}

\title{\large\textbf{Wrong and More Confident: A Field Experiment on Large Language Models Taking a Graduate Economics Exam}\thanks{Akimitsu: Department of Economics, State University of New York at Albany, econpi@icloud.com. The benchmark design, the grading rules, the analysis code, and the full result tables are public at \url{https://github.com/econpi/gerb-benchmark} and archived on Zenodo at \url{https://doi.org/10.5281/zenodo.21541073}. The sixty problem statements and their reference solutions are withheld to keep the benchmark out of training corpora and are available on request for research use under an agreement not to republish or train on them.}}
\author{Piyush Akimitsu\\ \textit{State University of New York at Albany}\\ \small \href{mailto:econpi@icloud.com}{econpi@icloud.com}}
\date{\today}

\begin{document}
\maketitle

\begin{abstract}
\noindent A \emph{red herring}, an irrelevant passage added to a problem, corrupts a language model's reasoning and, through it, its final answer, while the form of the response survives untouched. The benchmark, called the Graduate Economic Reasoning Benchmark (GERB), is sixty graduate-level microeconomics problems, each a detailed setup with a verified final answer and a step-by-step reference solution. Each problem has two versions, one with the red herring and one without, and each of those is asked in two ways, one requesting an explanation and one not. This is a within-subject $2\times2$ factorial experimental design. Thirty-eight language models answer all four versions of every problem. The clean problems (the control group) are already hard, with the models answering under sixty percent correctly on average. The red herring lowers the probability of a correct final answer by 12.3 percentage points, about a quarter of the models' mean accuracy of 0.525. The damage is largest on the problems the model rates as easy. Reasoning ability confers no protection, as the red herring's effect does not differ detectably across models with and without reasoning ability. It does change how the failure looks, since a model with no reasoning mode repeats one wrong answer across waves while a reasoning model wavers. The red herring also leads a model to rate a problem as easier than its clean version, while answering it wrong more often. Although open- and closed-weight models reach the same accuracy, the open-weight models reach it at a substantially lower cost per correct final answer. The form of the response is preserved even as its substance fails. The model still produces an explanation (explanation given), the final answer still follows from the reasoning shown (coherence), and, in the aggregate, it remains the same across waves (consistency).
\end{abstract}


\newpage
\doublespacing

\section{Introduction}

Language models report high accuracy on the benchmarks built to track progress toward general competence, including MMLU \citep{hendrycks2021mmlu} and GPQA \citep{rein2023gpqa}. Accuracy alone cannot settle what those scores mean. An accuracy figure conflates two sources. One is recall of a training pattern that closely matches the task. The other is construction of the answer from the mechanism operative in the task. Benchmarks near saturation do not separate the two, and saturation arrives alongside growing evidence of training-set contamination \citep{magar2022contamination, deng2024contamination, balloccu2024leak}. Making the benchmark harder, as with Humanity's Last Exam \citep{phan2025humanitys}, raises the ceiling but does not separate the two sources either. The source-of-accuracy question matters for safety and for deployment on work once done by expert reasoners \citep{eloundou2024labor, brynjolfsson2025generative}.

An instrument that can answer the question needs three things. The ground truth must be verifiable. The task must admit a within-task perturbation that leaves the content of the problem intact, so the comparison is identified from variation the leakage critique cannot reach. And the content must stay private while the design is published.

Economic reasoning is close to absent from the benchmarks that track this progress. GPQA covers biology, chemistry, and physics \citep{rein2023gpqa}, and the economics content of MMLU is undergraduate multiple choice \citep{hendrycks2021mmlu}. The economics benchmarks that do exist evaluate different objects, namely inference over pairs of economic events \citep{guo2024econnli}, the ordering of multi-event scenarios \citep{quan2024econlogicqa}, and decision-making by agents in unknown environments \citep{fish2025econevals}. To my knowledge, none asks a model to carry a stated economic model through a multi-step derivation to a single verified answer, which is what GERB does.

A microeconomics problem turns first on which of several qualitatively different situations holds, whether a constraint binds, whether the optimum is interior or at a corner, and whether the second-order conditions hold. That first judgment is discrete, and the arithmetic that follows is only as good as it. The domain is therefore a sharp test for a red herring, because an irrelevant passage does not have to break a calculation to change the answer. It only has to lead the reasoning astray before the arithmetic begins.

I built such a benchmark and call it the Graduate Economic Reasoning Benchmark (GERB). The domain is graduate-level analytical microeconomics. The corpus is sixty problems, each with one verified answer and a full worked solution. Each such problem is called a \pq{task}. Each task has two versions, one with a \emph{red herring}, an economically irrelevant passage, and one without, so the tasks with a red herring form the treated group and the tasks without one the control group. Each version is then asked in two ways, one requesting an explanation and one not. This is a within-task $2\times2$ factorial design. Thirty-eight language models answer all these four versions of all sixty tasks five times each, in independent stateless calls at temperature zero. A notable feature of these tasks is that even before any red herring is added, the corpus is hard, and on average the models answer under sixty percent of the clean problems correctly.

The headline result is that the red herring is costly since it reduces accuracy by about 12 pp. A loss of that size erases roughly a quarter of the correct answers. Asking the model to explain its answer improves accuracy. The two manipulations show no detectable interaction effect, and the red herring remains significantly harmful whether or not an explanation is requested. The damage is broad rather than the property of a weak tail, since the effect is statistically indistinguishable across the reasoning types.

On task versions requiring an explanation, the red herring lowers final-answer accuracy and reasoning correctness. The remaining features such as whether an explanation is given, coherence (whether the stated answer follows from the reasoning shown), and consistency (whether the answer stays stable across the five waves) are all unaffected. The models often get the red-herring tasks wrong yet, on average, rate them as easier than the clean versions.

This paper makes four contributions. (i) GERB, a within-task causal instrument for the source-of-accuracy question, built on sixty verifiable graduate-level microeconomics problems and a content-preserving perturbation, which yields a model-by-task panel instead of one number per model. It is released under a hybrid design, its content held private while its design is made public. \ifarxiv\ \footnote{The design, the grading rules, the analysis code, and the full result tables are public at \url{https://github.com/econpi/gerb-benchmark} and archived on Zenodo at \url{https://doi.org/10.5281/zenodo.21541073}, so that the benchmark is fully reproducible in method. The sixty problem statements and their reference solutions are withheld to keep the benchmark out of training corpora, and are available on request for research use under an agreement not to republish them or to train on them, so the test set can be applied to new models without entering a public crawl.} (ii) A causal estimate of the cost of the red herring, about twelve percentage points, with two-way clustered inference over sixty tasks and thirty-eight models. (iii) Evidence that the failure is a reasoning failure and not a failure of explanation, coherence, or consistency, together with a metacognitive inversion, in which the model rates the corrupted task as easier even though it answers it wrong more often. (iv) A verified per-model reasoning taxonomy over a forty-two-model catalog, with the provider-routing and truncation guards that keep long-reasoning models from being silently cut off. 

\section{Design and Data}\label{sec:design}

\paragraph{The tasks.} The instrument is a corpus of sixty microeconomics problems at graduate level. They span consumer choice, labor supply and time allocation, household production, monopoly and Cournot competition, production and factor demand, externalities and bargaining, price ceilings and floors, demand and supply elasticity, signaling and screening, and competitive equilibrium, and none is macroeconomics or econometrics.  Each problem states a structured multi-part setup, with named blocks for the environment, the budget or constraint, and the cost or payoff structure. Three properties make the corpus usable as an instrument. Each problem has exactly one objectively verifiable answer, so correctness is a fact rather than a matter of taste. The answers take four types, numeric, number-tuple, categorical, and symbolic. Each carries a full step-by-step reference solution whose intermediate quantities can be checked one by one, so a grader can check not only whether the final answer is correct but also whether the reasoning reached the answer through the relevant mechanism. And in their current form the tasks appear in no public benchmark and no indexable public source, so a model cannot have seen them as they are.\footnote{The problem setup and statements have a median of $226$ words, from $85$ at the shortest to $392$ at the longest, and the red herring adds a further $37$ words at the median. Solving one means carrying stated parameters through several steps of a model, not recognizing a pattern. The reference solutions are themselves a median of $238$ words of numbered steps, reaching $955$ at the longest.} 

\paragraph{The models.} The thirty-eight models come from a catalog of forty-two models whose default reasoning behavior I verify one by one against provider documentation, routing metadata, and an empirical probe. Type~1 reasons by default, Type~2 can reason but does not by default under these calls, and Type~3 has no reasoning mode. An independent judge grades every response, and a second judge regrades a sample to bound grader disagreement. The model under test never sees the answer or the worked solution.

\paragraph{The $2\times2$.} Each task is fielded in two versions, one with the red herring and one without, and each version is asked in two ways, one requesting an explanation and one not. The red herring $d\in\{0,1\}$ equals one when the economically irrelevant passage is added to the question text. It looks relevant and is not needed to solve the problem, and it leaves the verified final answer and the operative mechanism unchanged, so the treated and control versions of a task share the same final answer and the same reference solution. The explanation request $e\in\{0,1\}$ equals one when the prompt appends the verbatim sentence \pq{Provide a complete explanation.}, and the no-explain version appends nothing. Crossing the red herring with the explanation request gives four task versions of every task, two hundred and forty in all. Because the manipulation is within task, the red herring is orthogonal to task difficulty by construction, and task fixed effects absorb everything about the problem that is common to its four versions.

\paragraph{Execution.} Thirty-eight of the forty-two catalog models serve as contestants. The other four are the two graders and two held back on cost. Every model is reached through the OpenRouter gateway, which forwards each request to a provider that serves the model. Each contestant answers all two hundred and forty task versions in five waves at temperature zero. Every call is an independent, single-turn request with no shared state, sent with no reasoning parameter set, so each runs in the model's default deployed mode and carries nothing over from any other call. That is $60\times4\times38\times5=45{,}600$ planned responses.

\paragraph{Grading.} An independent judge model, xAI Grok~4.3, grades every response in its own conversation after the contestant has finished. The contestant never sees the verified answer or the worked solution, and only the judge does. On explanation-requested versions the judge returns four binary marks. The first is \emph{final answer correct}, one when the model's final answer matches the verified answer. The second is \emph{explanation given}, one when the response actually contains written reasoning. The third is \emph{reasoning correct}, one when the reasoning shown matches the reference worked solution. The fourth is \emph{coherence}, one when the final answer follows from the reasoning the model wrote. I reserve the word \emph{consistency} for the separate cross-wave measure.

On no-explanation versions the original pass graded only \emph{final answer correct}. I later regraded every one of those responses for \emph{explanation given}, because a model often explains even when the prompt asks only for an answer, and $58.3\%$ of them did. The last two marks, \emph{reasoning correct} and \emph{coherence}, are defined only when an explanation exists, so on no-explanation versions they can be graded only for the responses that volunteered one. Those responses are a self-selected group and not the whole condition, so I do not report a \emph{reasoning correct} or \emph{coherence} mean for the no-explanation arm, and I use the two marks only where the design holds the explanation factor fixed. Where an explanation was requested and none was returned, I code the response as failing \emph{reasoning correct} rather than leaving it ungraded.\footnote{A second, independent judge, Google Gemini~3.1~Pro, regraded a $2{,}750$-response subsample of the explanation-required responses to test whether the finding depends on who grades, and the two judges agree on $97.6\%$ of them. The full grading protocol, the model-access and routing details, and the inter-judge agreement table are in the appendix.}

\paragraph{The panel.} Of the $45{,}600$ planned responses, $45{,}402$ are scored. The missing $198$ are unscored because the response was truncated after retries ($133$) or the call errored after retries ($65$). They are spread evenly across the four conditions, and none is read as a wrong answer. The result is a model-by-task panel. A response is indexed by $i$, with model $m(i)$, task $q(i)$, and the two treatment indicators.

\paragraph{Specification.} I estimate linear probability models on that panel. The outcome $y_i\in\{0,1\}$ is final-answer correctness, one when the response's final answer matches the verified answer. The final answer is the objective and verifiable quantity the paper scores, so wherever the paper reports the probability of a correct answer it means a correct final answer, not the explanation or the reasoning around it. The saturated specification is
\begin{equation}
y_i \;=\; \beta_1 d_i + \beta_2 e_i + \beta_3\,(d_i e_i) + \gamma_{m(i)} + \alpha_{q(i)} + \varepsilon_i ,
\label{eq:main}
\end{equation}
where $d_i\in\{0,1\}$ is the red herring, equal to one when the irrelevant passage is present, $e_i\in\{0,1\}$ is the explanation request, equal to one when the prompt asks the model to explain, $\gamma_{m}$ are model fixed effects, and $\alpha_{q}$ are task fixed effects. There is one $\alpha_q$ per task, sixty in all, so the four task versions of a task share an intercept.  The task level is the finest level at which the design identifies the treatments. Both $d_i$ and $e_i$ are constant within a task version, so a task-version fixed effect would be collinear with them. There is no intercept, because one is not identified alongside the two sets of fixed effects. The parameter of interest is $\beta_1$, the effect of the red herring on the probability of a correct answer, averaged over the explanation factor when the interaction is absent. The coefficient $\beta_2$ is the effect of the explanation request and $\beta_3$ is the difference in differences. Standard errors are two-way cluster-robust, clustered by task and by model, over sixty tasks and thirty-eight models. The sample is the $45{,}402$ scored responses throughout. The per-model reasoning taxonomy is in the appendix.

\begin{figure}[!ht]
\centering
\includegraphics[width=\textwidth]{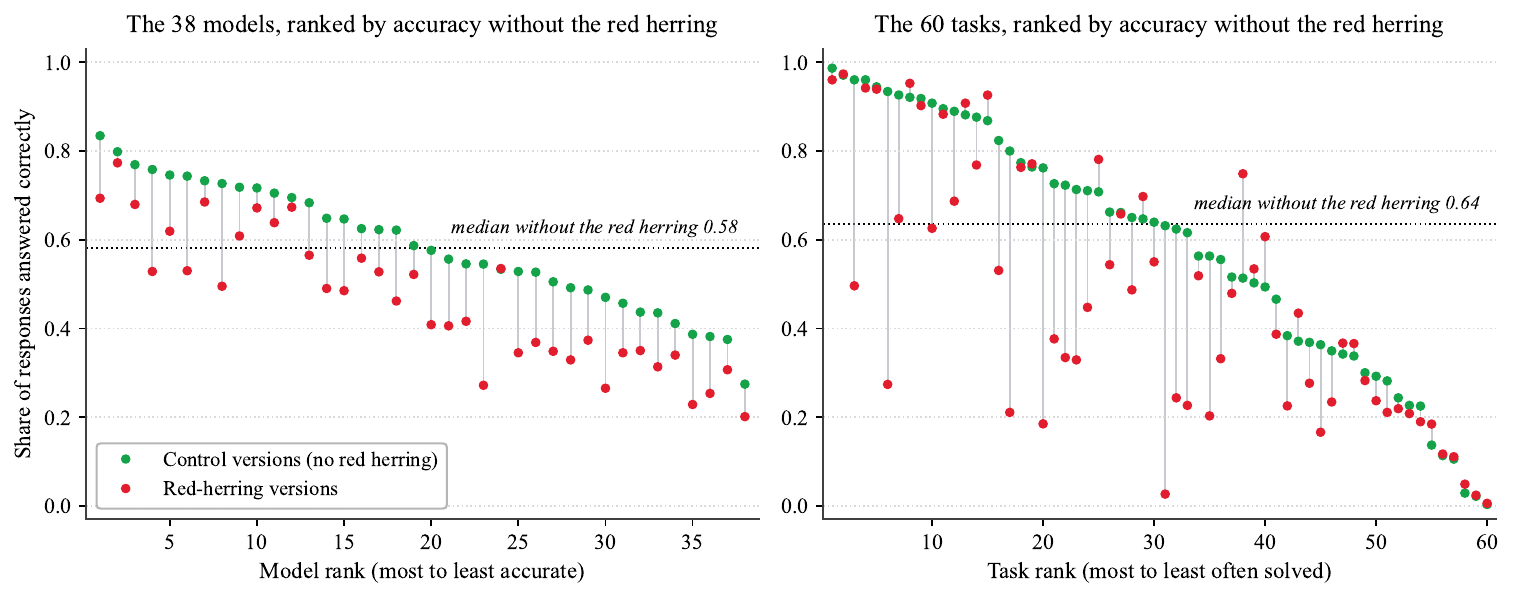}
\caption{How hard the corpus is, and what the red herring does to it. The left panel ranks the thirty-eight models by their accuracy on the control versions, and the right panel ranks the sixty tasks the same way. Green marks the control versions and red the red-herring versions, and the grey segment
is the drop. Accuracy is the share of scored responses whose final answer the judge marked
correct. No model is near ceiling on the control versions, and the tasks run from one answered
correctly on almost every attempt to one almost never answered correctly.}
\label{fig:accdist}
\end{figure}

Table~\ref{tab:supp_summary} reports accuracy and the four graded indicators for each of the four experimental conditions, over the $45{,}402$ scored responses.
\begin{table}[t]\centering\caption{Summary statistics by experimental condition. Four cells are deliberately left blank, and the note explains why.}\label{tab:supp_summary}
\begin{threeparttable}\resizebox{\linewidth}{!}{%
\begin{tabular}{@{}lcccccc@{}}\toprule
Condition & $N$ & Accuracy & Expl.\ given & Reas.\ correct & Coherent & Consistency \\\midrule
Control, no explanation & 11,344 & 0.552 & 0.612 &  &  & 0.860 \\
Red herring, no explanation & 11,358 & 0.436 & 0.553 &  &  & 0.867 \\
Control, explanation & 11,353 & 0.622 & 0.932 & 0.579 & 0.859 & 0.878 \\
Red herring, explanation & 11,347 & 0.491 & 0.924 & 0.459 & 0.846 & 0.869 \\\bottomrule\end{tabular}}
\begin{tablenotes}[flushleft]\scriptsize\item The four conditions are the $2\times2$ crossing of the red herring $d\in\{0,1\}$ with the explanation request $e\in\{0,1\}$. Accuracy is the share of scored responses whose final answer the judge marked correct, over every response in the condition. Mean accuracy across the $45{,}402$ scored responses is $0.525$.
\item \emph{Expl.\ given} is the share of responses that contain an explanation and is graded in all four conditions, over 22,477 graded responses on the two no-explanation conditions and all responses on the other two. On the no-explanation conditions the model was never asked to explain, so this column records how often it explained unprompted.
\item \emph{Reas.\ correct} and \emph{Coherent} are defined only for a response that contains an explanation. Where the explanation was requested they cover the whole condition, because a response returning no explanation is coded zero on reasoning correct. Where it was not requested, the only responses carrying these marks are the 13,095 in which the model explained unprompted. That subset is self-selected, and the red herring itself shrinks it from $0.612$ to $0.553$, so its mean would not be a condition mean and would not be comparable to the rows above. Those four cells are therefore left blank rather than filled with a selected average.
\item \emph{Consistency} indexes how stable the final answer is across the five waves, equal to $(5-a)/5$ where $a$ is the number of answer changes, so $1$ means the same answer on all five. It needs no explanation and is therefore filled in every condition, and it is a cell-level measure over the five waves rather than a per-response share. The red herring leaves it essentially unchanged.\end{tablenotes}\end{threeparttable}\end{table}

\section{Language Models Fail Two of Every Five Problems}\label{sec:baseline}

Before any perturbation is applied, the models answer more than four in ten of these problems incorrectly. Accuracy on the control versions is $0.587$. Figure~\ref{fig:accdist} shows that this is not produced by a weak tail dragging down an average. The strongest of the thirty-eight models answers $0.834$ of the control versions correctly, the median model $0.581$, and the weakest $0.274$. Thus, even the strongest contestant misses about one control problem in six. The right panel establishes a second property. Task accuracy on the control versions ranges from $0.987$ to $0.003$. A corpus that every model solves discriminates poorly, and so does one that no model solves. This corpus spans the range, which is why the estimate that follows is not an artifact of a ceiling.

\section{A Red Herring Lowers Accuracy by Twelve Percentage Points}\label{sec:main}

The explanation request raises the probability of a correct answer by $\beta_2 = 0.062$ (se $0.016$), the effect of asking for an explanation when no red herring is present. Holding the problem fixed, accuracy rises from $0.49$ to $0.56$, while the cost per correct answer rises only from $\$0.028$ to $\$0.031$. On accuracy alone the explanation is worth requesting. Whether it also offsets the red herring is a separate question that the interaction answers. Were a model required to set out its steps thereby led to identify the irrelevant passage and disregard it, the explanation would help more when the red herring is present than when it is absent. The interaction in column~(4) tests that difference. At $\beta_3 = -0.014$ (se $0.013$) it is indistinguishable from zero, so the benefit of the explanation does not differ detectably between the two arms and, equivalently, the red herring's loss does not differ detectably by whether an explanation was requested. The two manipulations are additive. Requesting an explanation raises the level of accuracy in both arms but does not narrow the gap the red herring opens.

The aggregate effect does not conceal heterogeneity in sign. Thirty-seven of the thirty-eight models are less accurate under the red herring. The single exception, Claude Sonnet~4, improves by $0.002$, a change well within its own confidence interval, and at the other extreme the most affected model loses $27.3$ percentage points. No model is measurably helped. The effect is a property of the class, not of a weak tail.\footnote{The per-model estimates and the forest plot are in Figure~\ref{fig:forest}. Each model's coefficient is estimated on its own responses alone, so the interval there is clustered by task only. With a single model there is no model dimension left to cluster on.}

\begin{table}[t]
\centering
\caption{Treatment effects on the probability of a correct answer.}
\label{tab:te_main}
\setlength{\tabcolsep}{3pt}
\begin{tabular}{@{}lcccc@{}}
\toprule
 & (1) & (2) & (3) & (4) \\
\midrule
Red herring $d$ & $-0.124^{***}$ & $-0.124^{***}$ & $-0.123^{***}$ & $-0.116^{***}$ \\
 & (0.027)        & (0.027)        & (0.027)        & (0.027)        \\
Explanation requested $e$ &      & $+0.063^{***}$ & $+0.062^{***}$ & $+0.069^{***}$ \\
 & & (0.016)        & (0.016)        & (0.019)        \\
$d \times e$ &                &                &                & $-0.014$       \\
                                 &                &                &                & (0.013)        \\
\midrule
Model FE & Yes & Yes & Yes & Yes \\
Task FE & No  & No  & Yes & Yes \\
Observations & 45{,}402 & 45{,}402 & 45{,}402 & 45{,}402 \\
$R^2$               & 0.094 & 0.098 & 0.375 & 0.375 \\
\bottomrule
\end{tabular}

\vspace{3pt}
{\footnotesize\singlespacing\setlength{\parindent}{0pt}\raggedright
The outcome is answer correctness $y_i\in\{0,1\}$, one if the response's final answer is correct. Each column is a linear probability model estimating eq.~(\ref{eq:main}). Column (1) has the red herring and model fixed effects, column (2) adds the explanation request, column (3) adds task fixed effects, and column (4) adds the interaction. Standard errors in parentheses are two-way clustered by task and model (60 tasks, 38 models).\par
\smallskip
Requesting an explanation ($e=1$) raises the share of responses that contain an explanation from $0.583$ to $0.928$, so $\beta_2$ measures the effect of asking for an explanation.\par
\smallskip
Stars: $^{*}p<0.1$, $^{**}p<0.05$, $^{***}p<0.01$.\par
}
\end{table}

\section{Reasoning Ability Confers No Protection}\label{sec:universality}

\begin{table}[t]
\centering
\caption{Mean accuracy by the model's reasoning type, with and without the red herring.
Reasoning raises the level of accuracy substantially but does not reduce the damage the
red herring does.}
\label{tab:reasontype}

\begin{threeparttable}
\setlength{\tabcolsep}{4pt}
\begin{tabular}{@{}lcccc@{}}
\toprule
 & & \multicolumn{2}{c}{Share answered correctly} & \\
\cmidrule(lr){3-4}
Reasoning type & Models & Control & Red herring & Difference \\
\midrule
Type 1 & 17 & 0.665 & 0.551 & $-0.114$ \\
Type 2 & 16 & 0.549 & 0.421 & $-0.128$ \\
Type 3 & 5 & 0.443 & 0.304 & $-0.139$ \\
\bottomrule
\end{tabular}
\begin{tablenotes}[flushleft]
\scriptsize
\item Each figure is the unadjusted share of scored responses the judge marked correct, pooling the explanation-requested and no-explanation versions. The fixed-effects estimates are in Table~\ref{tab:te_main} and the text. Reasoning type is defined in the surrounding text. Accuracy differs by type ($0.665$ for Type~1 versus $0.443$ for Type~3) while the red-herring fall is similar across types ($-0.114$ to $-0.139$).
\end{tablenotes}
\end{threeparttable}
\end{table}

Reasoning ability confers no protection. Models that reason by default lose as much accuracy to the red herring as models with no reasoning mode. The natural expectation is the opposite. If setting out the steps helped a model identify the irrelevant passage and set it aside, models that reason by default would be partly protected, and the effect would be smaller for them than for models that do not reason. Type~1 models reason by default, Type~2 can reason but do not under the calls made here, and Type~3 has no reasoning mode.\footnote{The type is a behavioral label, measured by whether a model emits a separate stream of \emph{reasoning tokens} before its visible answer under a default call, not a claim about whether it can reason. A Type~3 model still reasons, when it does, in its visible answer. OpenRouter's reasoning flag reports only whether a model accepts the reasoning parameter \citep{openrouter2025reasoning}, not whether it emits the stream. These numeric labels are not the Type~1 and Type~2 processes of dual-process theory \citep{evans2013dual, kahneman2011thinking}.}

The expectation is not borne out. Interacting the red herring with reasoning type, with the same fixed effects, clustering, and sample as Table~\ref{tab:te_main} and Type~1 as the reference, neither interaction is individually distinguishable from zero, $\delta_2 = -0.014$ (se $0.023$) for models that can reason but do not and $\delta_3 = -0.026$ (se $0.030$) for those with no reasoning mode, and the two are jointly insignificant ($F(2,37) = 0.40$, $p = 0.67$). A linear trend across the ordered types is also flat, at $-0.013$ (se $0.015$, $p = 0.37$). The full specification and the joint test are in Table~\ref{tab:typeinteraction}.

\begin{table}[htbp]\centering
\caption{The red-herring effect does not vary with reasoning type.}
\label{tab:typeinteraction}
\begin{threeparttable}
\begin{tabular}{@{}lc@{}}
\toprule
 & Answer correct $y_i$ \\
\midrule
Red herring $d$ (Type~1, reference) & $-0.114^{***}$ \\
 & $(0.029)$ \\
Red herring $\times$ Type~2 \ ($\delta_2$) & $-0.014^{}$ \\
 & $(0.023)$ \\
Red herring $\times$ Type~3 \ ($\delta_3$) & $-0.026^{}$ \\
 & $(0.030)$ \\
Explanation request $e$ & $+0.062^{***}$ \\
 & $(0.016)$ \\
\midrule
Model fixed effects & Yes \\
Task fixed effects & Yes \\
Observations & 45,402 \\
\midrule
Joint test $\delta_2=\delta_3=0$ & $F(2,37)=0.40$ \\
 & $p=0.671$ \\
Linear trend $d\times$ type & $-0.013\ (0.015)$ \\
 & $p=0.372$ \\
\bottomrule
\end{tabular}
\begin{tablenotes}[flushleft]\scriptsize
\item Linear probability model of answer correctness $y_i\in\{0,1\}$ on the red herring interacted with reasoning type, the explanation request, model fixed effects, and task fixed effects, over the 45,402 scored responses. Type~1 (reasons by default) is the reference, so the red-herring row is the Type-1 effect and $\delta_2$, $\delta_3$ are the additional red-herring effects for Type~2 (capable but off by default) and Type~3 (no reasoning mode). Standard errors in parentheses are two-way cluster-robust by task and model (60 tasks, 38 models). The joint test that both interactions are zero uses the same clustered covariance as the coefficients. The linear trend enters reasoning type as an ordinal $1/2/3$ variable interacted with the red herring. Stars: $^{***}p<0.01$, $^{**}p<0.05$, $^{*}p<0.1$.
\end{tablenotes}
\end{threeparttable}
\end{table}

This is a null result, not evidence that the types are alike. The intervals still admit modest differences in either direction, and if anything they point the other way, with the non-reasoning types slightly more harmed.\footnote{The per-model estimates show why no interaction separates the groups. Ordered by red-herring damage, the three types interleave across the whole range rather than forming separate bands, with models that reason by default among both the most and the least affected. Reasoning raises the overall level of accuracy, which the model fixed effects absorb. It does not change the slope with respect to the red herring.}

Figure~\ref{fig:forest} plots these per-model effects, ordered by harm and colored by reasoning type. The interval on each point is clustered one way by task, since with a single model there is no model dimension left to cluster on.

\begin{figure}[!ht]
\centering
\includegraphics[width=0.82\linewidth]{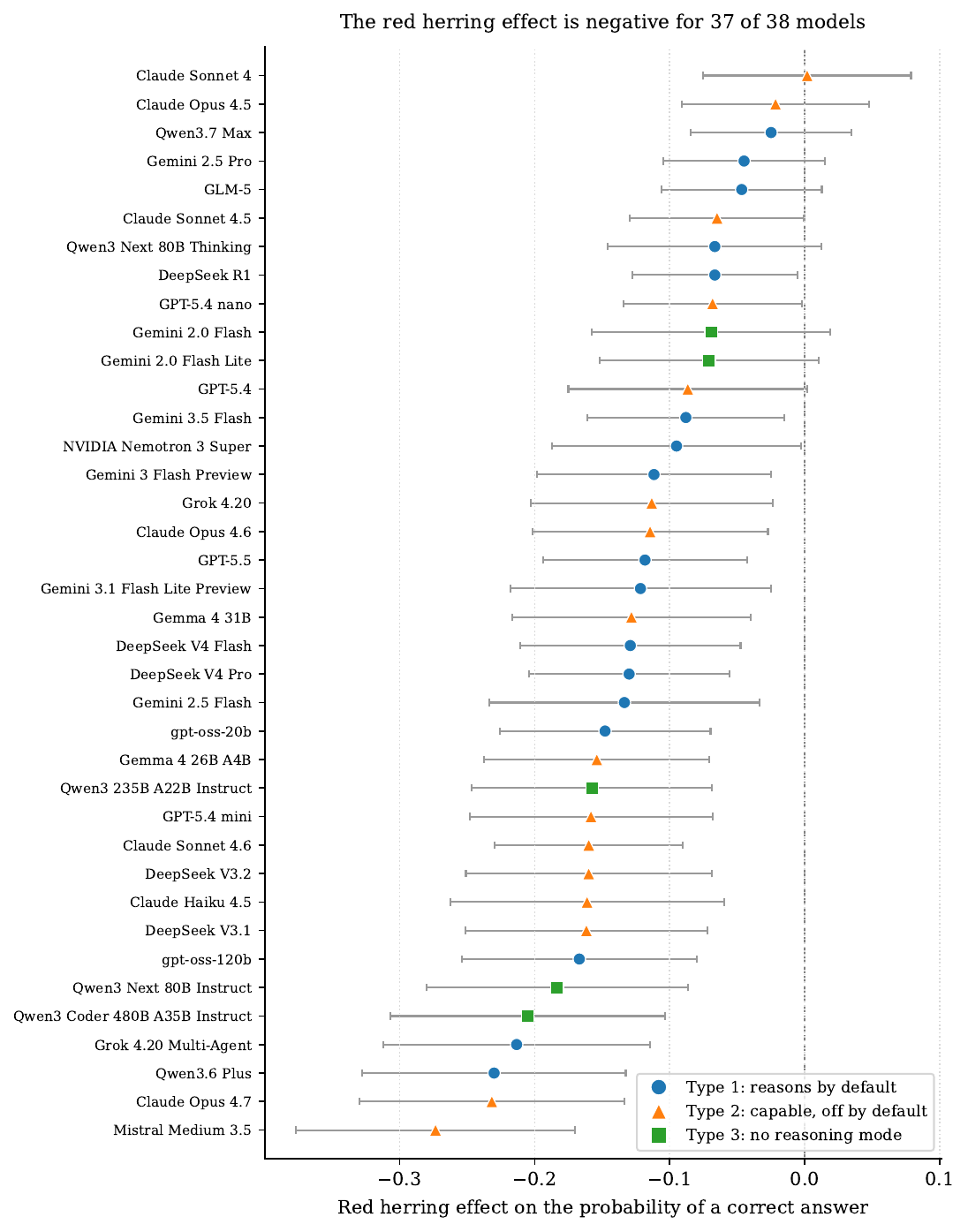}
\caption{The red-herring effect by model. Each point is one model's red-herring effect on the probability of a correct answer, the coefficient on the red herring in a per-model linear probability model of answer correctness on the red herring, the explanation request, and task fixed effects, estimated on that model's scored responses, with a 95\% confidence interval clustered by task. The interval is task-clustered because each point comes from a single model's regression. Models are sorted from most harmed to least harmed and colored by reasoning type. Thirty-seven of the thirty-eight are less accurate when the red herring is present.}
\label{fig:forest}
\end{figure}

Ranking models by how much the red herring moves them is not the same as ranking them by how often they are right, and the two orderings disagree. Figure~\ref{fig:accbymodel} shows each model's probability of a correct answer without the red herring, with it, and overall. Claude Sonnet 4 is the least moved model in Figure~\ref{fig:forest}, yet it sits at the sample mean here, because its answer barely changes rather than because it is correct. Qwen3.7 Max, the most accurate model at $0.79$, is also among the least moved. Robustness in the first figure can mean an unchanging mediocre answer as readily as a correct one.

\begin{figure}[!ht]
\centering
\includegraphics[width=0.72\linewidth]{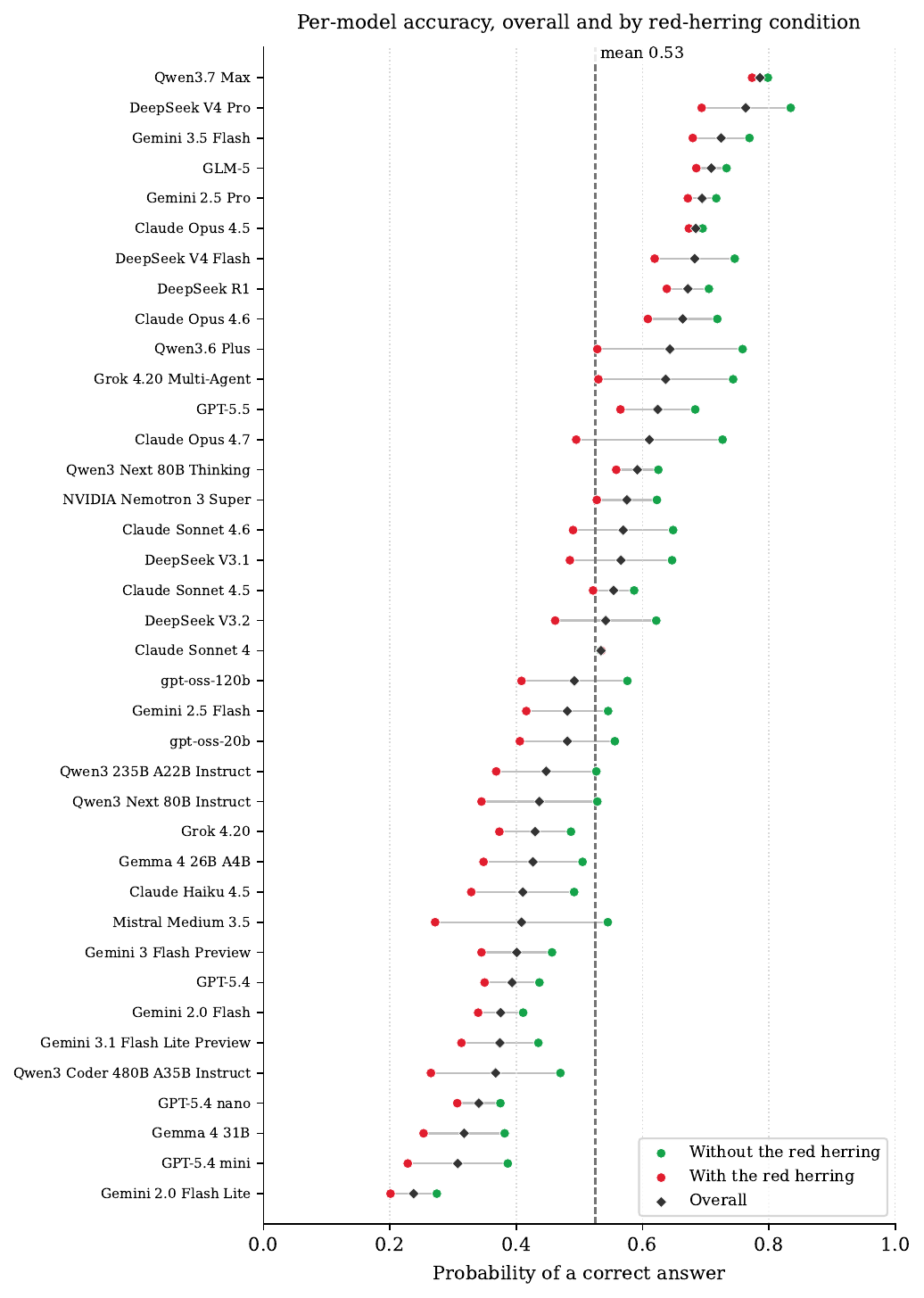}
\caption{Per-model accuracy, overall and by red-herring condition. For each model the probability of a correct answer without the red herring in blue, with the red herring in red, and overall as a grey diamond, sorted by overall accuracy. The dashed line is the pooled mean of $0.53$.}
\label{fig:accbymodel}
\end{figure}

Reasoning type does change one thing, and it is not accuracy. Table~\ref{tab:consist_bytype} re-estimates the interaction on two cell-level outcomes at once, answer correctness and a consistency score that records how often a model returns the same final answer across the five waves. The accuracy column reproduces the result above. The three types lose the same amount, and the joint test that both interactions are zero is far from rejection, $F(2,37) = 0.61$, $p = 0.55$. The consistency column does not behave this way. For a model that reasons by default the red herring lowers consistency by $0.015$ ($p = 0.02$), so its answer moves around more once the passage is inserted. For a model with no reasoning mode the interaction runs the other way and is larger, $+0.032$ ($p < 0.001$), enough to make the net effect positive. The joint test that reasoning type does not matter for consistency is rejected, $F(2,37) = 6.62$, $p = 0.003$, and the trend across the ordered types is monotone, $+0.018$ ($p = 0.002$). Pooling hides this. Averaged over all types the red herring barely moves consistency, which is what the next section reports, but that average is a cancellation of opposing type effects rather than a common null. A model with no reasoning mode does not merely get the corrupted task wrong more often. It gets it wrong more repeatably, returning the same diverted answer across waves, while a model that reasons wavers. Reasoning ability does not lower the accuracy cost of the red herring. It keeps the model from settling into a single wrong answer.

\begin{table}[htbp]\centering
\caption{Reasoning type does not change the accuracy the red herring costs, but it does change how repeatable the corrupted answer is.}
\label{tab:consist_bytype}
\begin{threeparttable}
\begin{tabular}{@{}lcc@{}}
\toprule
 & Answer correct & Consistency score \\
\midrule
Red herring $d$ (Type~1, reference) & $-0.114^{***}$ & $-0.015^{**}$ \\
 & $(0.029)$ & $(0.006)$ \\
Red herring $\times$ Type~2 \ ($\delta_2$) & $-0.014$ & $+0.024^{**}$ \\
 & $(0.023)$ & $(0.011)$ \\
Red herring $\times$ Type~3 \ ($\delta_3$) & $-0.030$ & $+0.032^{***}$ \\
 & $(0.027)$ & $(0.009)$ \\
Explanation request $e$ & $+0.062^{***}$ & $+0.009$ \\
 & $(0.017)$ & $(0.007)$ \\
\midrule
Model fixed effects & Yes & Yes \\
Task fixed effects & Yes & Yes \\
Cells & 9,027 & 9,027 \\
\midrule
Joint test $\delta_2=\delta_3=0$ & $F(2,37)=0.61$ & $F(2,37)=6.62$ \\
 & $p=0.546$ & $p=0.003$ \\
Linear trend $d\times$ type & $-0.015\ (0.014)$ & $+0.018\ (0.006)$ \\
 & $p=0.295$ & $p=0.002$ \\
\bottomrule
\end{tabular}
\begin{tablenotes}[flushleft]\scriptsize
\item Two linear probability models, one per column, estimated on the 9,027 cells, where a cell is one model on one task in one of the four conditions. Each regresses the cell outcome on the red herring interacted with reasoning type, the explanation request, model fixed effects, and task fixed effects, with Type~1 (reasons by default) the reference. The consistency score is the cell-level share recording how stable the final answer is across the five waves, equal to $1$ when the model returns the same answer on all five. Standard errors in parentheses are two-way cluster-robust by task and model (60 tasks, 38 models), and the joint test uses the same clustered covariance. The accuracy column reproduces the response-level Table~\ref{tab:typeinteraction} at the cell level. A positive consistency interaction means the red herring makes that type return the same answer more often, so a type that is also less accurate repeats a wrong answer more often. Stars: $^{***}p<0.01$, $^{**}p<0.05$, $^{*}p<0.1$.
\end{tablenotes}
\end{threeparttable}
\end{table}

\section{The Reasoning Is Corrupted While the Explanation, Coherence, and Consistency Are Preserved}
\label{sec:mech}

The accuracy drop established above shows that something was compromised, but not what. A model could fail by refusing to engage, by abandoning the problem and guessing, or by reasoning carefully in the wrong direction, and these call for very different responses from anyone deploying the system. Figure~\ref{fig:mech} separates them. The first four indicators are graded per response on the explanation-required half of the design. Each is regressed on the red herring alone, holding model and task fixed, so its coefficient is the effect of adding the irrelevant passage on that one feature. The fifth, consistency, is measured across the five waves.

\begin{figure}[!ht]
\centering
\includegraphics[width=\linewidth]{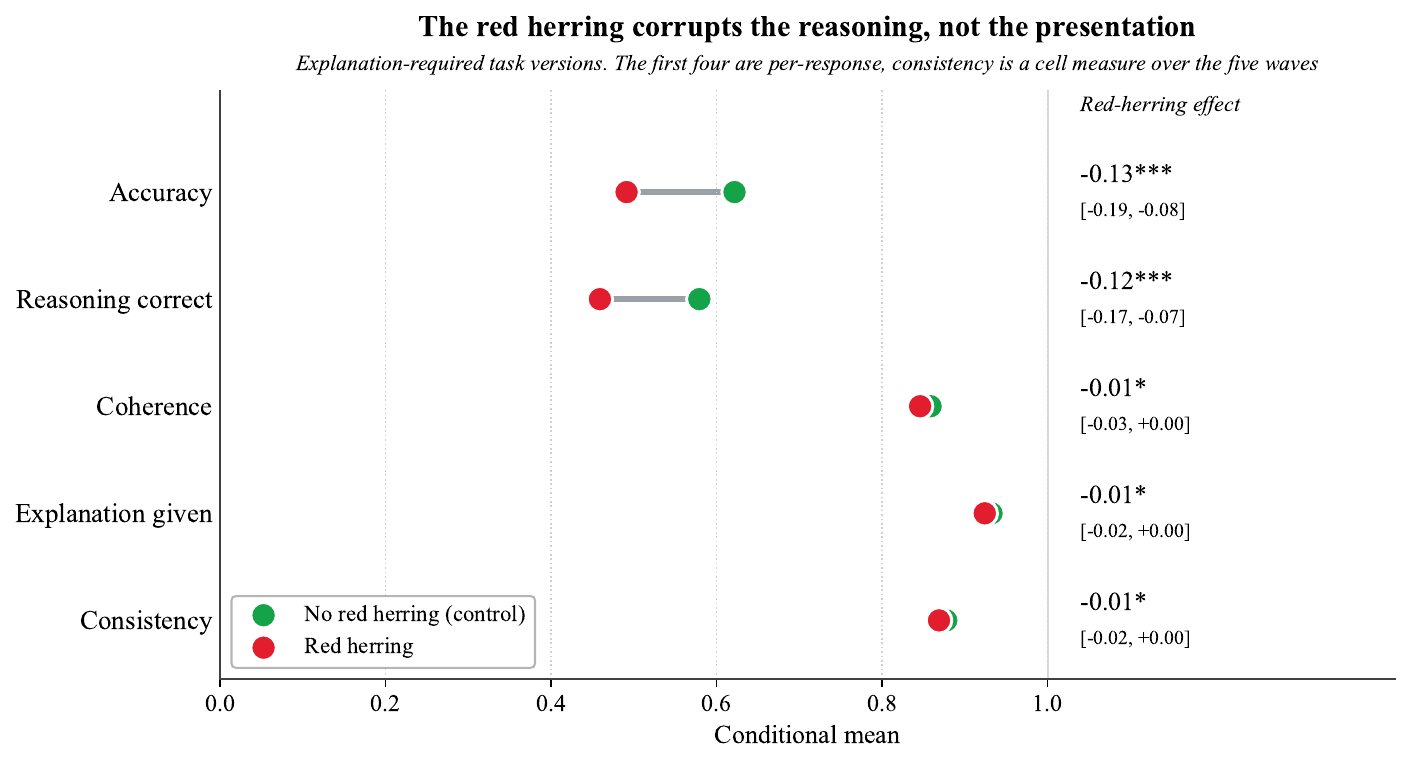}
\caption{Where the red herring does its damage. The first four rows are per-response zero-one indicators graded on the explanation-required task versions. The fifth, consistency, is a cell-level measure of whether the model gives the same answer across the five waves. Green marks the control mean and red the red-herring mean, both adjusted for model and task fixed effects, and the gap is the red-herring effect, reported with its 95\% confidence interval. Standard errors are two-way clustered by task and by model. Accuracy and correct reasoning fall sharply. Coherence, whether an explanation is given, and consistency barely move.}
\label{fig:mech}
\end{figure}

The red herring damages the substance of the reasoning and leaves the form of the response intact. It lowers the probability of a correct answer by thirteen percentage points, from $0.622$ in the clean arm to $0.491$ once the passage is added ($\beta_1^{\mathrm{acc}} = -0.130$, se $0.028$). It lowers the probability that the written reasoning is graded equivalent to the reference solution by almost as much, twelve percentage points, from $0.579$ to $0.459$ ($\beta_1^{\mathrm{reas}} = -0.120$, se $0.027$). The two losses are of a size because they are the same event. The red herring moves the model off the operative mechanism, and the incorrect answer is the downstream consequence of the incorrect reasoning.

The other three indicators are undisturbed. The probability that the model produces an explanation at all is $0.932$ in the clean arm and $0.924$ with the red herring, a decline too small to distinguish from zero ($\beta_1^{\mathrm{expl}} = -0.008$, se $0.004$). The probability that the final answer follows from the reasoning shown is $0.859$ against $0.846$, likewise indistinguishable from zero ($\beta_1^{\mathrm{coh}} = -0.013$, se $0.008$). Consistency across the five waves is similarly unaffected. On these versions the models are, as a rule, not refusing the task, not omitting the work, and not contradicting themselves. In the main they reason fluently and coherently toward an incorrect answer.

An incorrect answer under the red herring therefore does not, as a rule, reveal itself as wrong. It is delivered with a complete derivation, in the form of a correct response, because the model is faithfully reporting a line of reasoning that the red herring has diverted. Any check that reads the form of a response rather than its substance passes this failure through unaltered. Detection must come from outside the response, from a verified key, an independent solver, or a perturbation of the kind used here.

\section{The Damage Is Worst Where the Model Is Most Confident}\label{sec:difficulty}

The red herring does the most damage on the problems the model judges to be easy. A separate
elicitation asks each model to rate every version of every task from $1.0$ to $10.0$
for difficulty, in a fresh single-turn call that is independent of the answering protocol and that
instructs the model to rate the task without solving it. The rating of the control
no-explanation version is the \emph{baseline} rating $\delta_{qm}$, fixed before either treatment.
It lets me ask two questions the accuracy data alone cannot answer.

The models do not find this corpus easy in the first place. The mean baseline rating is $5.88$
with a standard deviation of $2.12$, and the ratings use the whole scale. The rating is
nonetheless a poor guide to whether the model will succeed. Its correlation with control accuracy
is $-0.185$, and Table~\ref{tab:diff_composition} shows the shape behind that weak association.
Accuracy on the control versions is flat near $0.67$ across the two easiest-rated quartiles,
declines to $0.594$ in the third, and falls to $0.381$ in the hardest. The models can therefore
identify the very hardest tasks as hard, but below that their sense of difficulty barely tracks
their own success rate.

\begin{table*}[t]
\centering
\caption{Accuracy with and without the red herring, by quartile of the model's own
baseline difficulty rating, together with the share of task-by-model pairs the model
never answered correctly in the control conditions.}
\label{tab:diff_composition}
\footnotesize
\begin{threeparttable}
\resizebox{\linewidth}{!}{%
\begin{tabular}{@{}llccccr@{}}
\toprule
 & Rating & \multicolumn{3}{c}{Share answered correctly} & Never solved & Task-by-model \\
\cmidrule(lr){3-5}
Quartile & range & Control & Red herring & Difference & in control & pairs \\
\midrule
Q1, rated easiest & 1.0--4.0 & 0.666 & 0.490 & $-0.176$ & 14.2\% & 565 \\
Q2 & 4.2--6.5 & 0.677 & 0.534 & $-0.143$ & 12.8\% & 617 \\
Q3 & 6.6--7.5 & 0.594 & 0.491 & $-0.102$ & 20.4\% & 604 \\
Q4, rated hardest & 7.6--9.5 & 0.381 & 0.320 & $-0.061$ & 42.3\% & 371 \\
\midrule
All four quartiles & 1.0--9.5 & 0.600 & 0.474 & $-0.126$ & 20.4\% & 2,157 \\
\bottomrule
\end{tabular}}
\begin{tablenotes}[flushleft]
\scriptsize
\item Quartiles are cut on the baseline rating (the model's own 1-to-10 rating of the control task, fixed before treatment) at the task-by-model pair level, roughly a quarter of the 2,157 rated pairs each. \emph{Control} and \emph{red herring} pool the two conditions without and with the red herring. Each figure is the share of scored responses the judge marked correct.
\item \emph{Never solved in control} is the share of pairs answered incorrectly on all ten control attempts (two control conditions by five waves). Such a pair cannot record a fall when the red herring is added, so it enters the difference as an arithmetic zero.
\end{tablenotes}
\end{threeparttable}
\end{table*}

Table~\ref{tab:diff_composition} also carries the column that governs how the quartile differences
should be read, namely the share of task-by-model pairs the model never answered correctly on any control
attempt, running from $14.2\%$ in the easiest-rated quartile to $42.3\%$ in the hardest. A pair the
model never solves cannot fall further when a red herring is added, so it enters the difference as
a zero by arithmetic. The hardest quartile therefore holds three times the share of pairs
incapable of registering damage.

\begin{table*}[t]
\centering
\caption{The average effect of the red herring on answer correctness, and how that effect
varies with the model's own baseline difficulty rating, estimated on the full sample and on
samples that exclude task-by-model pairs the model rarely solved without a red herring.}
\label{tab:diff_samples}
\footnotesize
\begin{threeparttable}
\setlength{\tabcolsep}{3pt}
\begin{tabular}{@{}lccrr@{}}
\toprule
 & Average effect of & Change in that effect & & Task-by- \\
 & the red herring & per rating point & & model \\
Sample & $\beta_1$ (SE) & $\beta_3$ (SE) & $N$ & pairs \\
\midrule
All pairs (the estimation sample) & $-0.1262$$^{***}$ (0.0274) & $+0.0178$$^{***}$ (0.0066) & 43,028 & 2,157 \\
Excluding never-solved pairs & $-0.1709$$^{***}$ (0.0322) & $+0.0135$$^{*}$ (0.0077) & 34,319 & 1,718 \\
Excluding pairs solved $<20\%$ & $-0.1848$$^{***}$ (0.0337) & $+0.0125$ (0.0083) & 32,292 & 1,616 \\
Excluding pairs solved $<50\%$ & $-0.2138$$^{***}$ (0.0366) & $+0.0090$ (0.0096) & 27,856 & 1,394 \\
\bottomrule
\end{tabular}
\begin{tablenotes}[flushleft]
\scriptsize
\item Outcome, fixed effects, and standard errors as in Table~\ref{tab:te_main}. $\beta_1$ is the average red-herring effect (eq.~\ref{eq:main}). $\beta_3$ is effect modification, the change in that effect per one added rating point (eq.~\ref{eq:gradient}, with the baseline rating $\delta^{c}_{qm}$ centered at its mean of 5.882), not a treatment effect. Each row restricts the sample by a task-by-model pair's control-side success rate, dropping all of that pair's responses in both arms. The first row is the full estimation sample.
\item Stars: $^{*}$ $p<0.1$, $^{**}$ $p<0.05$, $^{***}$ $p<0.01$.
\end{tablenotes}
\end{threeparttable}
\end{table*}

Table~\ref{tab:diff_samples} shows what that composition does. Dropping the pairs that could not
fall raises the average red-herring effect from $\beta_1 = -0.126$ (se $0.027$) to $-0.171$ (se
$0.032$), and dropping those solved less than half the time raises it to $-0.214$ (se $0.037$),
each estimated on the sample the restriction leaves. I keep the full sample as the headline. Excluding these pairs is
not a correction for a defect, since a model missing a hard task on every control attempt is an
ordinary outcome for a binary measure on hard material. The first row is the effect for the
population studied, and the later rows answer the narrower question of what happens among pairs that
could fall.

I turn to the gradient itself. I interact the red herring with the baseline rating, mean-centred so
the red-herring coefficient reads at average rated difficulty
\begin{equation}
y_i = \beta_1 d_i + \beta_2 e_i + \beta_3 (d_i \times \delta^{c}_{qm}) + \gamma\,\delta^{c}_{qm}
      + \gamma_m + \alpha_q + \varepsilon_i ,
\label{eq:gradient}
\end{equation}
with $y_i$ answer correctness, $\delta^{c}_{qm}$ the centred baseline rating, and the same model
and task fixed effects and two-way clustering as before, on the $43{,}028$ scored responses that
carry a rating. Every extra rating point of perceived difficulty shrinks the red herring's accuracy loss by
$\beta_3 = 0.018$ (se $0.007$), where $\beta_3$ measures effect modification, the change in the
red-herring effect per added rating point, and is not itself a treatment effect. Evaluated from
eq.~(\ref{eq:gradient}), the red-herring effect is $-0.195$ at a rating of $2$ and $-0.088$ at a
rating of $8$. The quartile differences in Table~\ref{tab:diff_composition} show the same pattern
without imposing the linear form, at $-0.176$ in the easiest-rated quartile against $-0.061$ in
the hardest, these being unadjusted differences in means rather than regression coefficients.

The mechanical objection is that accuracy is bounded below by zero, so there is simply less room to
fall on hard problems. One consideration addresses it. Control accuracy in the hardest-rated
quartile is $0.381$, which is nowhere near the floor. The qualification that does survive is compositional, the $42$ against
$14$ percent of pairs at exactly zero control accuracy noted above, and the robustness table reports the
restricted-sample and leave-out versions in full. The problems a model flags as routine are
therefore the ones the red herring derails, and also the ones a human reviewer is least likely to
check.

\begin{table}[t]
\centering
\caption{The red-herring effect is the same at the response level and the cell level.}
\label{tab:cellacc}
\setlength{\tabcolsep}{3pt}
\begin{tabular}{@{}lccr@{}}
\toprule
Outcome & $\beta_1$ & (se) & \multicolumn{1}{c}{$N$} \\
\midrule
Per-response $y_i\in\{0,1\}$ (headline) & $-0.1234^{***}$ & (0.0267) & 45{,}402 \\
Cell accuracy, mean over five waves & $-0.1227^{***}$ & (0.0267) & 9{,}113 \\
\bottomrule
\end{tabular}

\vspace{3pt}
{\scriptsize\setlength{\parindent}{0pt}\raggedright
$\beta_1$ is the red-herring coefficient in the headline specification, a linear probability model of answer correctness on the red herring, the explanation request, and model and task fixed effects, with standard errors two-way clustered by task and model. The first row uses every scored response. The second collapses the five waves in each model-task-version cell to one accuracy. Stars: $^{***}p<0.01$.\par}
\end{table}

\section{The Model Reads the Corrupted Task as Easier}\label{sec:firststage}

The model does not appear to notice that it has been misled. Asked only to judge how hard each task is, the models rate the versions carrying a red herring about two-tenths of a point \emph{lower} in difficulty, on average, than the clean versions of the same tasks, on the ten-point scale ($-0.181$, se $0.100$, $p = 0.071$), even though they answer them correctly less often. If the models recognized the red herring as an obstacle, they would rate those tasks harder, not easier. The estimate is only marginally significant, so the finding rests on the direction of the effect rather than its size.\footnote{The rating a model gives the same version it is answering is itself affected by the red herring, as just shown. Using it as a control would therefore soak up part of the very effect being estimated, so it never appears on the right-hand side of any regression in this paper. The separate difficulty elicitation exercise uses the baseline rating $\delta_{qm}$, the model's rating of the clean control version of the task, and hence does not move with red-herring or explanation request.}

\section{Open Weights Cost Less per Correct Answer}\label{sec:weights}

Accuracy and price are separate questions. I code each contestant on whether its trained weights
are released openly, following the NTIA availability test and checking every model against its
card. Table~\ref{tab:weightcost} shows the two groups reaching nearly the same accuracy at very
different prices, whether cost per correct answer is pooled or taken at the per-model median. The
gap does not reflect inexpensive weak models dragging down an average, since it survives
conditioning on control accuracy at every threshold, holding between $1.9$ and $3.4$ times. Open
weights do not, however, confer protection, since the red-herring effect does not robustly differ
by weight availability. Weight availability is not randomly assigned, so this describes the shape
of the market on offer today rather than an effect of releasing weights.
\begin{table}[t]\centering\caption{Cost per correct answer by weight availability, in US dollars.}\label{tab:weightcost}
\begin{threeparttable}\begin{tabular}{@{}lccrr@{}}\toprule
 & & & \multicolumn{2}{c}{Cost per correct (\$)} \\
\cmidrule(lr){4-5}
Weight & Models & Accuracy & Pooled & Median \\\midrule
Closed & 21 & 0.517 & 0.036 & 0.022 \\
Open & 16 & 0.530 & 0.014 & 0.003 \\\bottomrule\end{tabular}
\begin{tablenotes}[flushleft]\scriptsize\item Weight availability follows the NTIA test, whether the trained weights are released openly, verified per model card. Grok Multi-Agent is excluded as orchestration over a base model. Cost is the gateway-recorded per-call cost, so a reasoning model is charged for its reasoning stream. Pooled is total cost over total correct answers. Median is the median of the per-model ratios, lower because a few expensive closed models dominate the pooled figure. Above $0.5$ control accuracy the figures are $\$0.049$ closed and $\$0.015$ open. The supplement gives the threshold detail.\end{tablenotes}\end{threeparttable}\end{table}

\section{The Red Herring Lowers Accuracy Without Increasing Resource Use}\label{sec:resource}

A natural concern is that the red herring simply makes a model work harder, burning more tokens, dollars, and time on the task it gets wrong. Table~\ref{tab:resource} rules that out. Each column regresses a resource outcome on the red herring, the explanation request, their interaction, and model and task fixed effects, with standard errors two-way clustered by task and model. The token counts are estimated by PPML and the strictly positive cost and latency by log-OLS, so every coefficient is a proportional effect.

Pooled across the thirty-eight models, the red herring leaves completion tokens, reasoning tokens, and cost unchanged, and it lowers the time to a response by about twelve percent. The model does not deliberate longer over the task it gets wrong. If anything it commits faster, so the twelve-point accuracy loss arrives at no additional compute and would be invisible to anyone monitoring spend or token counts.

The reductions concentrate in the models that do not reason. Models that reason by default show no change in any resource. Type 2 answers about seventeen percent faster under the red herring, and Type 3 runs shorter, cheaper, and faster still, though with only five Type 3 models its estimates are imprecise. These models can produce long and costly explanations when asked, as the explanation-request coefficients show, yet the red herring pushes them toward a quicker and cheaper wrong answer rather than a more effortful one. The one manipulation that raises resources is the explanation request, and it is cheap for the models that already reason and expensive for those that do not, so buying back accuracy through explanation costs least exactly where reasoning is already on.

Across models, accuracy rises with resource use but not with the model's own sense of difficulty. Figure~\ref{fig:accresource} plots each model's accuracy against its mean cost, latency, completion tokens, and perceived difficulty. Accuracy rises with the three resources, at Spearman rank correlations of $0.76$, $0.68$, and $0.76$, so the more expensive, slower, and more verbose models are the more accurate ones. Against perceived difficulty the relationship is flat, at $-0.07$, so a model's own rating of how hard the tasks are barely tracks whether it gets them right. The red herring does not move a model along the resource frontier. It leaves cost and tokens unchanged and only shortens the response, so the accuracy it destroys is not bought back by spending more.

\begin{figure}[!ht]
\centering
\includegraphics[width=\linewidth]{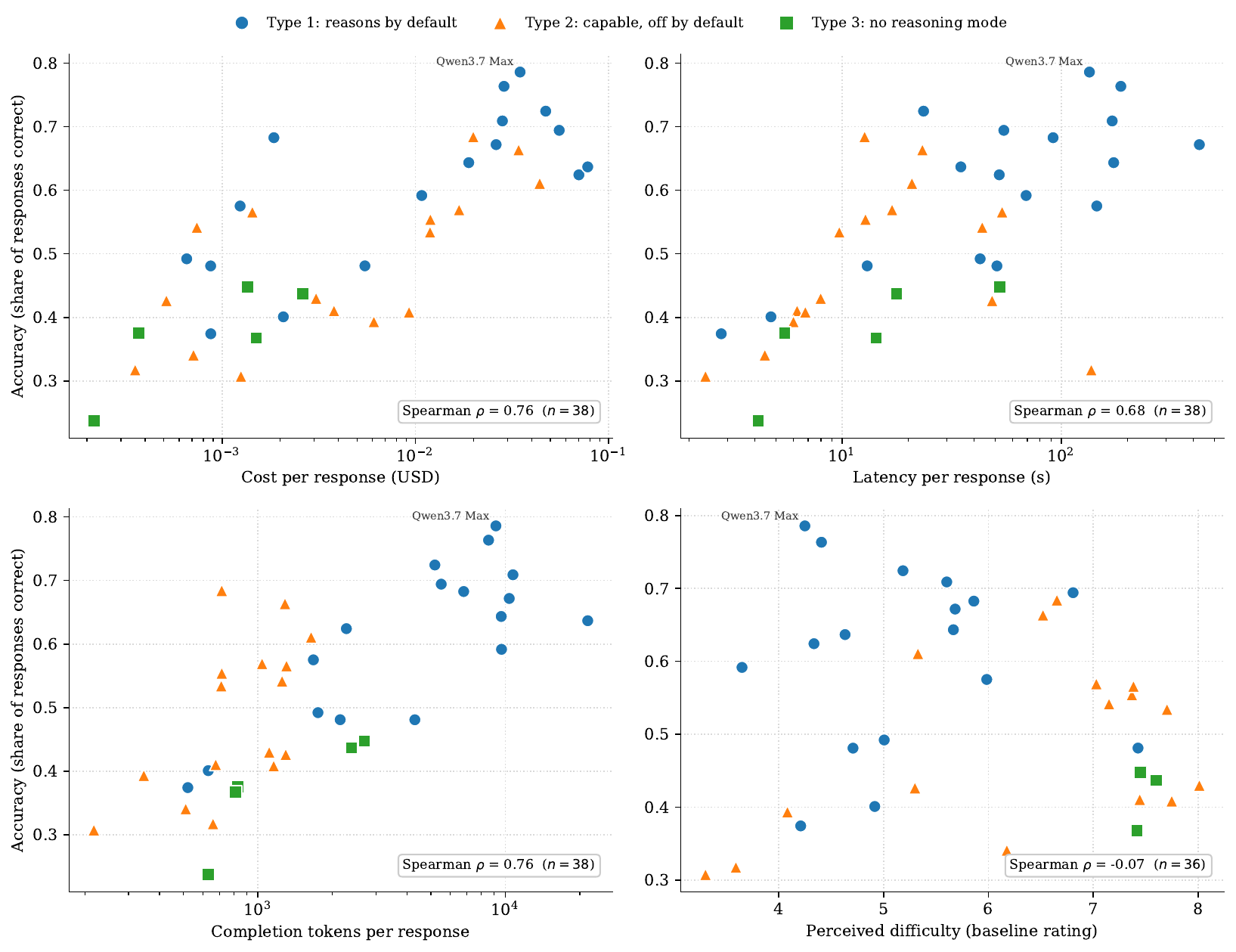}
\caption{Accuracy against resource use and perceived difficulty, one point per model. Each panel plots a model's accuracy, the share of its scored responses the judge marked correct, against its mean cost, latency, completion tokens, and perceived difficulty (the mean baseline rating). The three resource axes are logarithmic. Points are both colored and shaped by reasoning type, Type~1 circles, Type~2 triangles, and Type~3 squares. Accuracy rises with the three resources, at Spearman rank correlations of $0.76$, $0.68$, and $0.76$, but is flat against perceived difficulty, at $-0.07$. The red herring lowers accuracy without raising any resource, so the loss is not recovered by spending more.}
\label{fig:accresource}
\end{figure}

\begin{table}[p]\centering
\caption{Effect of the red herring and the explanation request on resource use, by reasoning type.}
\label{tab:resource}
\begin{threeparttable}
\begin{tabular}{@{}lcccc@{}}
\toprule
 & Completion & Reasoning & Cost & Latency \\
 & tokens & tokens & (USD) & (s) \\
Estimator & PPML & PPML & Log-OLS & Log-OLS \\
\midrule
\addlinespace
\multicolumn{5}{@{}l}{\textit{Panel A. Pooled, all 38 models}} \\[1pt]
\quad Red herring $d$ \ ($\beta_1$) & $-0.026$ & $-0.012$ & $-0.044$ & $-0.127^{***}$ \\
 & $(0.041)$ & $(0.048)$ & $(0.033)$ & $(0.038)$ \\
\quad Explanation request $e$ \ ($\beta_2$) & $+0.169^{***}$ & $+0.030$ & $+0.679^{***}$ & $+0.548^{***}$ \\
 & $(0.052)$ & $(0.026)$ & $(0.091)$ & $(0.078)$ \\
\quad $d\times e$ \ ($\beta_3$) & $+0.004$ & $+0.010$ & $+0.025$ & $+0.100^{***}$ \\
 & $(0.024)$ & $(0.022)$ & $(0.025)$ & $(0.031)$ \\
\quad Observations & 45,155 & 16,702 & 45,118 & 45,402 \\
\midrule
\multicolumn{5}{@{}l}{\textit{Panel B. Type 1, reasons by default (17 models)}} \\[1pt]
\quad Red herring $d$ \ ($\beta_1$) & $-0.018$ & $-0.012$ & $+0.026$ & $-0.037$ \\
 & $(0.045)$ & $(0.048)$ & $(0.038)$ & $(0.041)$ \\
\quad Explanation request $e$ \ ($\beta_2$) & $+0.092^{**}$ & $+0.030$ & $+0.514^{***}$ & $+0.308^{***}$ \\
 & $(0.039)$ & $(0.026)$ & $(0.127)$ & $(0.067)$ \\
\quad $d\times e$ \ ($\beta_3$) & $+0.009$ & $+0.010$ & $-0.028$ & $+0.041$ \\
 & $(0.025)$ & $(0.022)$ & $(0.029)$ & $(0.029)$ \\
\quad Observations & 20,285 & 16,702 & 20,248 & 20,305 \\
\midrule
\multicolumn{5}{@{}l}{\textit{Panel C. Type 2, capable but off by default (16 models)}} \\[1pt]
\quad Red herring $d$ \ ($\beta_1$) & $-0.058$ & --- & $-0.075^{*}$ & $-0.182^{***}$ \\
 & $(0.044)$ &  & $(0.041)$ & $(0.055)$ \\
\quad Explanation request $e$ \ ($\beta_2$) & $+0.497^{***}$ & --- & $+0.726^{***}$ & $+0.671^{***}$ \\
 & $(0.088)$ &  & $(0.113)$ & $(0.103)$ \\
\quad $d\times e$ \ ($\beta_3$) & $-0.011$ & --- & $+0.047$ & $+0.138^{***}$ \\
 & $(0.048)$ &  & $(0.031)$ & $(0.049)$ \\
\quad Observations & 19,055 & --- & 19,055 & 19,185 \\
\midrule
\multicolumn{5}{@{}l}{\textit{Panel D. Type 3, no reasoning mode (5 models)}} \\[1pt]
\quad Red herring $d$ \ ($\beta_1$) & $-0.144^{***}$ & --- & $-0.193^{**}$ & $-0.259^{***}$ \\
 & $(0.035)$ &  & $(0.084)$ & $(0.082)$ \\
\quad Explanation request $e$ \ ($\beta_2$) & $+0.697^{***}$ & --- & $+1.105^{***}$ & $+0.969^{***}$ \\
 & $(0.161)$ &  & $(0.245)$ & $(0.294)$ \\
\quad $d\times e$ \ ($\beta_3$) & $+0.049$ & --- & $+0.133$ & $+0.180^{*}$ \\
 & $(0.065)$ &  & $(0.100)$ & $(0.094)$ \\
\quad Observations & 5,815 & --- & 5,815 & 5,912 \\
\bottomrule
\end{tabular}
\begin{tablenotes}[flushleft]\scriptsize
\item Each cell is a regression of the resource outcome on the red herring $d$, the explanation request $e$, their interaction $d\times e$, model fixed effects, and task fixed effects, on the sample named in each panel, with standard errors two-way cluster-robust by task and model. Columns 1--2 (token counts) are PPML (\texttt{ppmlhdfe}); columns 3--4 (cost and latency, continuous and strictly positive) are OLS on the natural log (\texttt{reghdfe}). Coefficients are proportional, so $\beta$ is a $100\,(e^{\beta}-1)\%$ effect (PPML) or approximately a $100\,\beta\%$ effect (log-OLS). The log-cost regressions drop $37$ zero-cost responses. Reasoning tokens are blank for Types~2 and~3, which do not reason by default, so PPML separates them out and the pooled reasoning column is the Type~1 estimate. Type~3 has only five model clusters, so its clustered standard errors are unreliable. Stars: $^{***}p<0.01$, $^{**}p<0.05$, $^{*}p<0.1$.
\end{tablenotes}
\end{threeparttable}
\end{table}

\section{Robustness}\label{sec:robust}

The red-herring effect survives every check I run. Across specifications the coefficient stays between $-0.10$ and $-0.21$ and is significant at the $1\%$ level, whether I change the fixed-effect partition, restrict to complete cells or drop the truncation-heavy tasks, split judge-graded from rule-graded responses, or cluster one way rather than two. Two checks are worth stating here. The red-herring coefficient is larger, not smaller, on the $7{,}434$ responses graded by a deterministic rule with no judge call, at $\beta_1 = -0.210$ (se $0.053$) against $-0.103$ (se $0.024$) under the judge, which is the opposite of what a lenient or confused judge would produce. And a wild cluster bootstrap over the sixty task clusters gives $p<0.001$, so the result does not rest on the two-way asymptotics with only thirty-eight model clusters.

A model's answer to a fixed prompt is a stochastic draw rather than a deterministic function, so a single call is a noisy measurement of the model's success on a task version. The five waves are repeated measurements of the same cell, which estimate that success probability far more precisely than one call and make the design reproducible against sampling noise. Collapsing each cell to its mean accuracy over the five waves and re-estimating the headline specification leaves the red-herring effect unchanged, as Table~\ref{tab:cellacc} shows. Table~\ref{tab:supp_robust} reports the full battery, every specification, sample, grader, and clustering scheme I run, and the coefficient stays between $-0.10$ and $-0.21$ and significant at the $1\%$ level throughout.

\begin{table}[t]\centering\caption{Full robustness of the red-herring effect on answer correctness. Each row re-estimates the red-herring coefficient $\beta_1$ under a different specification, sample, grader, or clustering scheme.}\label{tab:supp_robust}
\begin{threeparttable}\setlength{\tabcolsep}{2pt}\resizebox{\linewidth}{!}{%
\begin{tabular}{@{}lcccr@{}}\toprule
Specification & $\beta_1$ & (se) & $p$ & $N$ \\\midrule
Baseline (model FE only, two-way cluster by task and model) & $-0.1235$ & $(0.0267)$ & $<0.001$ & 45,402 \\
Adding task fixed effects (headline equation (1)) & $-0.1234$ & $(0.0267)$ & $<0.001$ & 45,402 \\
Task $\times$ explanation-arm fixed effects (120) & $-0.1234$ & $(0.0267)$ & $<0.001$ & 45,402 \\
Complete model x task pairs only (no missing cell) & $-0.1245$ & $(0.0268)$ & $<0.001$ & 44,480 \\
Drop truncation-heavy tasks Q07, Q08 & $-0.1278$ & $(0.0274)$ & $<0.001$ & 43,972 \\
Judge-graded cells only & $-0.1033$ & $(0.0242)$ & $<0.001$ & 37,968 \\
Rule-graded (deterministic) cells only & $-0.2099$ & $(0.0525)$ & $<0.001$ & 7,434 \\
Clustering by task only (one-way) & $-0.1235$ & $(0.0256)$ & $<0.001$ & 45,402 \\
Cell accuracy, mean over five waves (model and task FE) & $-0.1227$ & $(0.0267)$ & $<0.001$ & 9,113 \\
Wild cluster bootstrap 95\% CI (task clusters) & $[-0.176,\,-0.076]$ & --- & $<0.001$ & 45,402 \\\bottomrule\end{tabular}}
\begin{tablenotes}[flushleft]\scriptsize\item Outcome, answer correctness $y_i\in\{0,1\}$. Unless noted, model fixed effects are included and standard errors are two-way cluster-robust by task and model (60 tasks, 38 models). The wild cluster bootstrap does not re-estimate the coefficient, which stays at the $-0.1235$ of the one-way row, and reports instead the $95\%$ confidence interval shown for $\beta_1$ together with the $p$-value, from resampling the 60 task clusters with 999 Rademacher draws. The estimate stays between $-0.10$ and $-0.21$ and is significant at the $1\%$ level throughout.\end{tablenotes}\end{threeparttable}\end{table}

\section{Related Work and Limitations}\label{sec:related}

\paragraph{Contamination and memorization.} Public benchmarks such as MMLU \citep{hendrycks2021mmlu} and GPQA \citep{rein2023gpqa} saturate quickly, and the tasks increasingly leak into the training corpora of the models they score \citep{magar2022contamination, deng2024contamination, balloccu2024leak}. A headline accuracy figure then mixes reasoning ability with recall. Benchmark audits reach the same place from the quality side. \citet{reuel2024betterbench} score twenty-four widely used benchmarks against forty-six criteria and find most report no statistical significance. Theory adds a strategic channel. \citet{truong2026strategic} model the benchmark as a principal-agent contract in which the developer rationally specializes in the measured slice. The hybrid design I adopt, private content and public design, follows from the contamination logic above rather than from any external prescription. My identification sidesteps contamination rather than testing for it. Both arms of the comparison use the same task, so anything the model memorized about that task is differenced out.

\paragraph{Irrelevant context and perturbation.} That irrelevant text degrades model accuracy is well
established. \citet{shi2023distracted} show grade-school arithmetic accuracy dropping sharply when one
irrelevant sentence is appended, and \citet{mirzadeh2024gsm} report losses up to sixty-five percent
from a single irrelevant clause added to GSM8K problems \citep{cobbe2021gsm8k}. Closest to the present
design, \citet{naeini2023herrings} find that models fixate on red-herring cues in creative-reasoning
puzzles. Further work catalogs perturbation sensitivities across a robustness benchmark
\citep{wang2024rupbench} and traces compositional failures as reasoning depth grows
\citep{dziri2023faith}. My contribution is not the phenomenon but the design and the domain. The red
herring is manipulated within task on graduate-level problems that leave the operative mechanism and
the verified answer untouched, so its effect is a causal estimate net of model and task fixed effects
rather than a descriptive accuracy drop on public, largely elementary benchmarks.

\paragraph{Reasoning models and automated grading.} Chain-of-thought prompting \citep{wei2022chain} and self-consistency sampling \citep{wang2023selfconsistency} raise accuracy by making reasoning explicit. The current generation instead hides the chain on a separate billed channel, a convention begun by the o1 family \citep{openai2024o1} and adopted by DeepSeek-R1 \citep{guo2025deepseekr1}. Whether that channel is active by default is a model-level fact, not a prompt setting. Hence, I verify it for every model rather than trusting reasoning flags. For grading I follow the LLM-as-judge protocol of \citet{zheng2023judging} and add a second independent judge. The model under test is never in the judge pool.

\paragraph{Experiments on language models.} A growing literature experiments on language models to a
different end, treating the model as a stand-in for a human subject and asking whether it reproduces or
predicts human behavior, from simulated economic agents \citep{horton2023llms} to predicting human
field-experiment results from model responses \citep{chen2025predicting}. In my design the task is the
experimental subject that receives the perturbation, presented under every condition and serving as its
own control, while the model is the agent whose reasoning is under evaluation. What is measured is the
model's own competence, not its fidelity to a person. This within-subject design follows the framework
of \citet{list2025within}.

\paragraph{Metacognition and calibration.} A separate literature asks whether a model knows how well it
is doing. Language models are often miscalibrated and overconfident on the very material they answer
incorrectly \citep{steyvers2025know, steyvers2025metacognition}, and they lack the metacognitive
monitoring that reliable reasoning demands \citep{griot2025metacognition}. My metacognitive inversion
adds a causal contrast to these correlational findings. Adding the red herring lowers accuracy and, at
the same time, lowers the model's own difficulty rating: a controlled corruption drives the model's
confidence in the wrong direction.

\paragraph{Limitations.} Three caveats are to be noted. First, the domain is analytical microeconomics. It is untested whether the red herring bites as hard in law, medicine, or code. Second, models are as deployed through a router rather than as local weights. This places a provider's serving stack and quantization inside the estimate \citep{lamalfa2024lmaas}. Third, model behavior behind a fixed API name drifts \citep{chen2024chatgpt}, so the accuracy levels are specific to the run window but the estimated effect is not. This is confirmed by the cross-wave consistency for each model-task version.

\section{Conclusion}\label{sec:conclusion}
A benchmark score runs two things together, recalling a familiar pattern and constructing an answer
from the mechanism a problem is about. A within-task red herring separates them. It leaves the
problem, its verified answer, and its worked solution unchanged, yet it lowers accuracy by about
twelve percentage points, and it does so whether or not a model reasons by default and whether or
not its weights are released openly. The failure is not a refusal and not a matter of form. The
model still produces a complete explanation, gives an answer consistent with the reasoning it
shows, and rates the corrupted problem, if anything, as easier than the clean one. What degrades is
the reasoning itself. Reasoning ability changes how a model fails, not how often. Without a reasoning mode a model settles on a single wrong answer under the red herring while a reasoning model wavers, so the accuracy cost is shared but the error is more entrenched.

The finding has two implications. For measurement, a headline accuracy figure is not evidence of robust reasoning, and a cheap content-preserving perturbation reveals how much of that figure rests on the wording of a task. For deployment, the failure is silent. It comes with a full and coherent derivation and with greater confidence rather than less, so nothing within the response marks it as wrong. Detection must come from outside, from a verified answer or an independent solver. A final observation concerns cost. Open-weight and closed-weight models reach the same accuracy on these problems, but the open-weight models do so at a markedly lower cost per correct answer, so on this task the choice between them turns on price rather than reasoning ability.

\clearpage
\begin{spacing}{1.0}
\bibliographystyle{aea}
\bibliography{AGI_ML_Paper}
\end{spacing}

\newpage
\section{Appendix}

\subsection{Measurement Notes on Model Behavior}

This section records technical findings from the pilot runs. They shaped the measurement choices for the main study. I report them so that the results can be read and reproduced with care.

Each model has a limit on how many tokens\footnote{A token is the unit of text a model reads and writes, roughly a short word or a piece of a word.} it can return in one response. I set this limit to eight thousand tokens. Most models finish well below it. One model does not. Gemini 3.1 Pro produces a long internal reasoning stream before its answer, and for this model that stream counts against the limit. On three of six pilot calls it reached the limit and stopped before the answer was complete. The reported stop reason\footnote{Each call reports why the model stopped. A natural stop, reported as \cw{stop}, means the model finished its answer. A length stop, reported as \cw{length}, means the model reached the output cap and was cut off before finishing, so the response is incomplete and cannot be scored.} was \cw{length} and not a natural stop. The response was cut off and could not be scored. Models that also reason at length are not all affected in the same way. Grok 4.20 Multi-Agent returned more than twenty thousand tokens on some calls and still finished on its own, because its reasoning tokens do not count against the same limit. A single fixed limit therefore acts differently across models. A low limit does not save money. A model is billed only for the tokens it generates, so a model that finishes in seven hundred tokens costs the same under a high limit as under a low one. A low limit only adds the risk that a verbose model is cut off, which wastes the call and returns nothing usable. For the full study I raise the limit to thirty two thousand tokens. This is high enough that no model is cut off, and it adds cost only for the few models that truly need the room. I check the stop reason on every call and treat any value other than a natural stop as a failed cell to be rerun.

Some models place their reasoning in a hidden channel and return only a short final answer. The judge sees the visible answer and not the hidden reasoning. On the tasks that ask for an explanation, GPT-5.5 often returned a bare answer, such as a single word or a single number, and gave no visible explanation, while still spending several hundred tokens of hidden reasoning. Under my scoring this counts as no explanation. This is a deliberate choice. The task asks for an explanation. A model that is asked for an explanation and returns none has not met the request, whatever it did in private. I treat the visible answer as the object of measurement throughout.

I treat whether a model reasons as a variable, but I define it with care. My model list carries a label that marks a model as reasoning capable. I assigned that label when I built the list, so it reflects my reading and not a statement from the provider. The slugs\footnote{A slug is the short identifier a model is called by, such as \cw{deepseek/deepseek-r1}.} and the prices come from the live model service, but the label is my own. The pilot shows that the label is a weak guide to behavior. Several models that I marked as reasoning capable returned no hidden reasoning tokens at all under a plain call, while a few others returned many thousands. The label says what a model can do. It does not say what the model did on the task. I also checked the provider's own list of supported parameters\footnote{This is a field that OpenRouter, the gateway I route every call through, publishes for each model. OpenRouter forwards each call to one of the independent providers that host the model. The field lists the parameters a model accepts, not the settings it uses by default. See \url{https://openrouter.ai/docs/guides/best-practices/reasoning-tokens} for the gateway's reasoning parameter conventions.}, in case it gave a cleaner label. It does not. It records whether a model accepts a reasoning parameter, and almost every current model does, including models such as Claude Sonnet and Gemma that do not reason on their own. So that list marks whether reasoning can be requested, not whether a model reasons, and it does not split the set in a useful way.

I therefore prefer a measured variable. For each model I record whether it returns hidden reasoning tokens, and I form a model level indicator from this. The indicator marks whether the model uses a separate hidden reasoning channel under my calls. This is not the same as whether the model reasoned at all. A model can reason in plain view by writing out its steps and still return no hidden tokens. The Claude Opus models do this. So the indicator measures the use of a hidden channel, not the presence of reasoning, and I name it on those terms. I use the indicator at the model level and not at the level of a single call. The number of hidden tokens on one call is a response to that task, so it sits after the treatment and would be a bad control. The clean question is whether the red herring effect differs between models that use a hidden channel and models that do not. That comparison is the interaction\footnote{An interaction term lets one effect differ across groups. Its coefficient here is the difference in the red herring effect between models that use a hidden channel and models that do not.} of the indicator with the red herring, and it survives model fixed effects\footnote{A model fixed effect is a separate intercept for each model. It absorbs everything about a model that does not change across tasks, including its overall accuracy, so the comparison of interest is made within each model and not between models.} because the red herring varies within each model. The hidden token count itself stays in the data as a descriptive measure of effort, not as a control. I call every model in its default mode and do not set the reasoning parameter\footnote{Many models accept an optional setting that turns on a longer internal reasoning pass. Leaving the setting unset is the model's default, so the behavior I record is each model's default and the call is identical across models.} that most models accept. The behavior I record is therefore the model's default, and the call is the same for every model.

The price of a model and the length of its answers are not related in a simple way. A high price per token does not imply a long answer and a low price does not imply a short one. In the pilot the Claude Opus models carry a high price per token yet returned about seven hundred tokens per answer, so each call was cheap. Grok 4.20 Multi-Agent carries a low price per token yet returned more than ten thousand tokens per answer on average, so each call was not cheap. For this reason I measure the cost of each model on the real tasks rather than read it from the headline price. A small pilot on a single task with a few repeats gives the per model cost that I use to plan the full run.

I set the sampling temperature to zero on every call. A temperature of zero does not make a model perfectly repeatable. Reasoning models change their final answer across repeats of the same task. DeepSeek R1 changed its answer on about one cell in five. A model without a long reasoning stream is steadier but not perfect. Gemini 2.0 Flash repeated its answer on most cells but not all. I therefore run each cell five times and call each run a wave. From the five runs I build the accuracy and consistency measures, which capture both how often the model is correct and how steady its answer is across runs.

I grade short answers in two ways. When the answer is a short token that matches the key in an exact and safe way I mark it by rule. When the answer is longer or open to reading I send it to the judge. In my checks the rule never marks a wrong answer as correct. This keeps grading fast on the easy cases and reserves the judge for the cases that need judgement.

I run many calls at once. A worker is one parallel line of execution that handles one cell from start to finish. Running several workers at once shortens the wall clock time of a run, since a reasoning model can take minutes on a single call. The number of workers does not change any result. Each cell is a separate request at temperature zero, so a cell decided by one worker is identical to the same cell decided alone. Writes to the output files are guarded by a lock so that parallel workers never collide.

I track spending as the run proceeds and stop at a hard cap. The provider returns the dollar cost of each call. When that figure is missing or zero I fall back to a price computed from the token counts and the published per token rates. The fallback is an upper bound because it ignores any discount, so the cap can never undercount. I store the cost of the model call and the cost of the judge call as separate fields and keep their sum, so the price of each cell can be read directly. Because the run uses many workers, the cap can be passed by a small amount equal to the calls already in flight when it trips, and I set the cap with room for this.

For each call I store a full record. It holds the input tokens, the output tokens, the hidden reasoning tokens when the provider reports them, the total tokens, the time taken, the stop reason, the dollar cost, and the model that actually served the request. This record lets me audit cost and speed for every model and lets a reader reproduce the accounting.

\subsection{Provider Routing and Output Caps}

OpenRouter does not serve the models itself. It forwards each call to one of several independent providers that host the model, and it spreads calls across them.\footnote{OpenRouter's provider routing documentation states verbatim that ``by default, requests are load balanced across the top providers to maximize uptime,'' and that a caller can override this by adding a \cw{provider} object to the request body. See \url{https://openrouter.ai/docs/guides/routing/provider-selection}.} The providers that serve one model do not share a single output cap. Each provider sets its own maximum on the number of tokens a single response may contain. For most models in the study every provider allows at least the thirty two thousand tokens I request, so the choice of provider does not matter. For a minority it does. DeepSeek R1 is the clearest case. One provider that serves it caps a response near four thousand tokens while another allows near sixteen thousand. R1 reasons by default and its reasoning counts against the cap, so a call routed to the low capacity provider is cut off in the middle of the reasoning and returns an incomplete answer. The stop reason is a length stop rather than a natural stop. On the hardest items the loss is systematic, since those are the items on which the model reasons longest, so the bias would fall on the very treatment effect I am trying to measure.

I fix this at the routing layer. For every model I queried OpenRouter's per model endpoint list and read the output cap of each provider.\footnote{The endpoint list is a JSON resource at \texttt{https://openrouter.ai/api/v1/models/<slug>/endpoints} for any model slug. It returns one record per endpoint with a \cw{max\_completion\_tokens} field that gives the output cap. I read it for every model on 27 May 2026. The DeepSeek R1 listing, for example, is at \url{https://openrouter.ai/api/v1/models/deepseek/deepseek-r1/endpoints} and returns two endpoints, one with \cw{max\_completion\_tokens} equal to four thousand ninety six and one equal to sixteen thousand.} I then exclude, for each model, every provider whose cap is below the thirty two thousand tokens I request, but only when the model has another provider with a higher confirmed cap. This rule never drops a model's best provider, and it never acts on a cap reported as zero or missing, which the service uses for unknown rather than for low. Table~\ref{tab:caps} lists the twelve models for which at least one provider is excluded and names the excluded providers. NVIDIA Nemotron 3 Super is a special case. No provider serves it above about sixteen thousand tokens, so there is no higher capacity provider to route to. I exclude none for it and rely on the stop reason check instead.

I implement the exclusion as a blacklist rather than a whitelist, and the choice matters.\footnote{A blacklist names the providers to avoid and lets the service choose among the rest. A whitelist names the only providers that may be used. A blacklist can only narrow the pool, so a stale or misspelled name simply excludes nothing and the call still goes through. A whitelist that matches no provider leaves the service with no endpoint, and the call fails. OpenRouter's documented fields realize both. The \cw{ignore} field is a list of provider slugs to skip for a request, the \cw{only} field is a list of provider slugs to allow, the \cw{order} field is a list of provider slugs to try in order, and the boolean \cw{allow\_fallbacks} defaults to true, and the routing rule I use sets only \cw{ignore}. See \url{https://openrouter.ai/docs/guides/routing/provider-selection} for the field definitions. The documentation also states that a base provider slug, for example \cw{google-vertex}, matches all endpoints for that provider including any regional variants, which is why the routing rule uses base slugs.} An earlier attempt used an ordered whitelist that pinned each model to a named provider and forbade any fallback. It failed. The service could not match the provider tags on this account and returned no endpoints, so the call died. The blacklist avoids this. It can only remove providers from the pool, so a name that no longer matches removes nothing and the call still completes through some other provider. The routing change is therefore fail safe. It can improve the provider choice but it cannot send a call to a dead end.

The routing rule lowers the risk of a cut off answer but does not by itself prove that none remains, so I keep a second guard at the level of the response. Every call reports its stop reason. When a response ends with a length stop I retry it up to twice, since the service may route the retry to a higher capacity provider. If it still ends with a length stop I mark the cell as truncated and never record it as a score. A truncated cell is rerun and is never read as a wrong answer. This guard runs on every call and catches any case the routing rule leaves, including a model such as Nemotron whose best provider still caps below my request. In the validation runs the two guards together left no truncated cell. A full pass across all thirty-eight contestants returned no length stop.

\begin{table}[t]
\centering
\caption{The twelve models for which I exclude at least one low capacity provider from routing. Providers are named by the base slug used in the routing rule. Built on 27 May 2026 from OpenRouter's per model endpoints interface.}
\label{tab:caps}

\begin{threeparttable}
\begin{tabular}{@{}lp{7.5cm}@{}}
\toprule
Model & Providers excluded for a low output cap (base slug) \\
\midrule
DeepSeek R1 & \texttt{azure} \\
DeepSeek V3.1 & \texttt{sambanova} \\
DeepSeek V3.2 & \texttt{deepinfra}, \texttt{sambanova} \\
DeepSeek V4 Pro & \texttt{deepinfra} \\
DeepSeek V4 Flash & \texttt{deepinfra} \\
gpt-oss-20b & \texttt{siliconflow} \\
gpt-oss-120b & \texttt{siliconflow} \\
Qwen3 Next 80B Instruct & \texttt{deepinfra} \\
Qwen3 235B Instruct-2507 & \texttt{deepinfra}, \texttt{google-vertex}, \texttt{novita}, \texttt{streamlake} \\
GLM-5 & \texttt{deepinfra}, \texttt{venice} \\
Gemma 4 26B A4B & \texttt{deepinfra}, \texttt{venice} \\
Gemma 4 31B & \texttt{ambient}, \texttt{deepinfra}, \texttt{venice} \\
\bottomrule
\end{tabular}
\begin{tablenotes}[flushleft]
\footnotesize
\item Rule. For each model I exclude a provider when every endpoint of that provider caps output below the thirty two thousand tokens I request and the model has another provider with a higher confirmed cap. The rule never drops a model's best provider and never acts on a cap reported as zero or missing. DeepSeek R1 keeps Novita, which allows about sixteen thousand tokens, and drops only Azure, which caps near four thousand. NVIDIA Nemotron 3 Super has no provider above about sixteen thousand tokens, so it is not listed, and the stop reason guard catches any answer that exceeds its cap. The rule is keyed by the model slug in the run script.
\end{tablenotes}
\end{threeparttable}
\end{table}

\subsection{Independence and Statelessness of API Calls}

A premise of the empirical design is that each cell decision is an independent observation. The premise has two parts. The first is that the model under test sees only the prompt the runner constructed for that cell and nothing else from outside that prompt. The second is that no prior cell, prior question, or prior wave leaves a residue that influences the current call. Both parts are guaranteed by the Chat Completions contract that OpenRouter implements.\footnote{The endpoint reference is at \url{https://openrouter.ai/docs/api/api-reference/chat/send-chat-completion-request} and accepts the \cw{messages} array as the sole carrier of conversational context. A \cw{session\_id} field is supported but its documented purpose is request grouping for routing and observability, not server-side conversation memory \citep{openrouter_apiref_2026}.} Each request is processed in isolation, the gateway retains no state across requests, and the caller is responsible for assembling whatever conversational context the request requires.

The runner reflects this contract directly. For a causal cell the runner constructs a \cw{messages} array with one user message containing the question, extended in the explanation condition with the explanation prompt. No \cw{system} message is included. The array of any cell is constructed from that cell's prompt alone. No prior cell, prior question, or prior wave is reused, referenced, or carried over.

Two consequences for the inference follow. First, the no-carryover condition that \citet{list2025within} requires for a credible within-subject design is satisfied at the protocol level, not assumed. Second, the analytic panel can be read as a collection of independent draws at the cell, model, and wave grain, with the model fixed effect absorbing whatever the model brings to the call from its training and weights but not from any prior cell of this study.

\subsection{Temperature, Stochasticity, and Provider-Side Normalization}

The runner sets the sampling temperature to zero on every call. The intent is to make the model's output as close to repeatable as the model allows, so that wave-to-wave variation reflects the model's residual stochasticity rather than deliberately injected sampling jitter. The choice does not bind uniformly across models, and the validity of the wave-level consistency measure depends on understanding which models honor temperature and which do not.

For the open-weight families that I call without a reasoning parameter (Qwen Instruct, DeepSeek V3.1 and V3.2 in the \cw{deepseek-chat} face, Gemma 4, Mistral Medium 3.5) and for the closed-weight chat models that do not internally reason (Anthropic Claude in its non-thinking face, Gemini 2.0 Flash and Flash-Lite), temperature is honored at the provider's sampler. For these models temperature zero produces the closest the provider can manage to a fixed point of the decoding distribution.

For the OpenAI reasoning families, the o-series and the GPT-5 family in all of its variants, the temperature parameter is not supported. Microsoft's documentation for Azure OpenAI lists \cw{temperature}, \cw{top\_p}, \cw{presence\_penalty}, \cw{frequency\_penalty}, and several other sampling controls as unsupported with reasoning models \citep{azure_openai_reasoning_2026}.\footnote{See \url{https://learn.microsoft.com/en-us/azure/foundry/openai/how-to/reasoning}, which states ``The following are currently unsupported with reasoning models: \cw{temperature}, \cw{top\_p}, \cw{presence\_penalty}, \cw{frequency\_penalty}, \cw{logprobs}, \cw{top\_logprobs}, \cw{logit\_bias}, \cw{max\_tokens}.''} On a direct call to OpenAI's API such a request returns HTTP 400. Through OpenRouter the same call succeeds. The gateway silently normalizes the unsupported parameter and the request is accepted. I verified this empirically on 2026-06-17 by sending GPT-5.5 a chat completion request with \cw{temperature = 1.5}, a value that is both out of the conventional range and unsupported for this model family. OpenRouter returned HTTP 200, the model produced a normal answer of six output tokens, and twenty-nine reasoning tokens were billed. No error was raised. The substantive implication is that temperature does not bind for the OpenAI reasoning family in my pipeline. Sampling proceeds with the model's internal default. The verification script and the recorded payloads and responses are in the accompanying \texttt{VERIFICATION\_CITATIONS.md}.

For DeepSeek R1 the situation differs. DeepSeek's reasoning model documentation states that ``setting \cw{temperature}, \cw{top\_p}, \cw{presence\_penalty}, \cw{frequency\_penalty} will not trigger an error but will also have no effect'' \citep{deepseek_reasoning_2026}.\footnote{See \url{https://api-docs.deepseek.com/guides/reasoning_model}, section on the reasoning model API contract.} I verified this empirically on 2026-06-17 by sending the same reasoning prompt to R1 twice with temperature zero. The two visible final answers were both correct and identical in value but the visible derivations differed token by token. The reasoning-token counts differed by roughly thirteen percent and the two responses ended at slightly different points within the maximum-tokens window. This is the behavior the documentation predicts and is the same behavior the wave-level consistency measurements pick up.

The validity-relevant conclusion is that wave-level consistency is a measurement of the model's residual stochasticity under my intended setting. For families that honor temperature, consistency below one means the sampler did not return to the same point of its temperature-zero distribution across waves. For families that do not honor temperature, consistency below one means the model's internal default sampling produced different sequences. Both readings are economically meaningful. In either case the model has not committed to a single answer under repeated exposure to the same prompt, and that is the property the consistency outcome captures \citep{guo2025deepseekr1}.

\subsection{API Access and What the Contestant Model Can See}

The contestant model on any given call has access to exactly what the runner places in the request body. This section describes what the body contains, what it explicitly omits, and what the deployed model can therefore reach at the provider's end.

The body of a contestant call contains five fields and at most one optional sixth. The five required fields are \cw{model}, \cw{messages}, \cw{temperature}, \cw{max\_tokens}, and a \cw{usage} options object that asks the gateway to include the cost report in the response. The optional sixth is \cw{provider}, set only for the twelve models in the routing rule of Table~\ref{tab:caps}. The body does not contain a \cw{system} message, a \cw{tools} array, a \cw{response\_format} directive, a \cw{search\_parameters} object, or a \cw{thinking} or \cw{reasoning} field. The absence of these fields determines what the model can reach during the call.

For the Anthropic and OpenAI families the absence of a \cw{tools} array means no tool is exposed. For the Google Gemini family the absence of a \cw{tools} array means that grounding with Google Search is not enabled. Google's grounding documentation requires the caller to attach a \cw{google\_search} tool to the request's \cw{tools} field for grounding to occur \citep{gemini_grounding_2026}.\footnote{See \url{https://ai.google.dev/gemini-api/docs/google-search}, which presents the required configuration as \cw{Tool(google\_search=GoogleSearch())} inserted into the \cw{tools} list.} I verified this empirically on 2026-06-17 by asking my judge model, Gemini 3.1 Pro, for a breaking news story from the past twenty-four hours and instructing it to emit a fixed refusal phrase if it could not access live data. The model emitted the refusal phrase. The model also billed one hundred eighty-eight reasoning tokens on the one-sentence question, which is consistent with its Type 1 classification as a mandatory-reasoning model.

For the xAI Grok family the documented default is the opposite. xAI's chat completion reference states that the \cw{search\_parameters.mode} field defaults to \cw{on}. The documentation reads, ``\cw{on (default)}: the model will search in every sources for relevant data'' \citep{xai_apiref_2026}.\footnote{The canonical URL is \url{https://docs.x.ai/developers/rest-api-reference/inference/chat}, the request reference for the chat completion endpoint. The path \url{https://docs.x.ai/docs/api-reference} also serves the same content via permanent redirect.} A naive reading is that the Grok calls in this study were doing live web and X search by default, since the runner sends no \cw{search\_parameters} object. I tested this empirically on 2026-06-17 by sending three Grok 4.20 requests through OpenRouter to the same prompt about a real-time topic, one with no \cw{search\_parameters} field, one with \cw{search\_parameters.mode = off} set explicitly, and one with \cw{search\_parameters.mode = on} set explicitly. All three returned the same fixed refusal phrase the prompt asked for in the absence of live data. The documented xAI default does not propagate through OpenRouter for this model. The gateway either strips or normalizes the field, and the Grok contestants in my run did not access live X or web data during the call. I disclose both facts honestly. The upstream xAI default is on by documentation. The realized behavior through my gateway is off by empirical test. The same caveat applies to Grok 4.20 Multi-Agent and to Grok 4.3 in the judge role. The judge's grading does not require search in any case because the rubric and the reference answer are provided in the judge's own prompt.

The open-weight families in the contestant pool, DeepSeek (V3.x, R1, and V4), Qwen (Instruct, Thinking, Coder, Max, and Plus), Gemma 4, gpt-oss-20b and 120b, GLM-5, Mistral Medium 3.5, and NVIDIA Nemotron 3 Super, have no native tool access at the weights level. There is nothing for an absent \cw{tools} field to disable.

Beyond tools and grounding, the body of the call is what the model sees. The runner does not send a \cw{system} message and the gateway does not synthesize one. Whatever default system content a provider may apply at its own boundary is part of the model-as-deployed and is not under my control.

A final disclosure relevant to the open-weight contestants concerns precision. Several providers in the OpenRouter pool serve open-weight models at reduced precision (FP8, FP6, INT8, INT4) for cost and throughput. DeepInfra documents this explicitly. Its documentation states, ``Changing precision changes the numbers your model sees and produces, so it inevitably changes behavior'' \citep{deepinfra_quantization_2025}.\footnote{See \url{https://deepinfra.com/blog/precision-to-quantization-faster-cheaper-llms}, which flags long chains of thought and code-and-math tasks as the regimes where precision matters most.} The routing rule of Table~\ref{tab:caps} excludes the providers I identified as low-output-cap. For the open-weight contestants that remain, the \cw{native\_model} column records the provider that actually served each cell, and the per-call cost in the audit log lets a future analyst verify the routing for each row.

Table~\ref{tab:apiverify} summarizes the empirical verifications of the API-behavior claims relevant to this study. Each row pairs a claim drawn from provider documentation with a terminal test run against the same OpenRouter endpoint used by the runner. The tests were authored as standalone Python scripts and run from the same shell that drove the en-masse data collection. Total cost across all eight calls was \$0.0086. The raw script source and the captured responses are preserved in the supplementary file \texttt{VERIFICATION\_CITATIONS.md} accompanying the analysis code.

\begin{table}[t]
\centering
\caption{Empirical verifications of API behavior on the OpenRouter \cw{/v1/chat/completions} endpoint, run on 17 June 2026. Each row records one terminal test against a documentation claim. The full call log is in the supplementary verification file.}
\label{tab:apiverify}
\scriptsize
\setlength{\tabcolsep}{4pt}
\renewcommand{\arraystretch}{1.15}
\begin{threeparttable}
\begin{tabular}{@{}>{\raggedright\arraybackslash}p{2.9cm}>{\raggedright\arraybackslash}p{4.3cm}>{\raggedright\arraybackslash}p{4.3cm}>{\raggedright\arraybackslash}p{3.9cm}@{}}
\toprule
test (model) & claim verified & method & result \\
\midrule
Grok 4.20\tnote{a} (3 calls) & xAI's documented default \cw{search\_parameters.mode = on} does not propagate through OpenRouter & Three calls to the same prompt about a real-time topic: (i) no \cw{search\_parameters} field, (ii) explicit \cw{mode = off}, (iii) explicit \cw{mode = on} & All three returned the same prompted refusal phrase. Live search was not active in any of the three. \\
GPT-5.5\tnote{a} & OpenAI reasoning models reject \cw{temperature}, and OpenRouter normalizes it before forwarding & One call with \cw{temperature = 1.5}, an out-of-range and provider-unsupported value & HTTP 200, normal six-token answer, 29 reasoning tokens billed. No error raised. \\
DeepSeek R1\tnote{a} & DeepSeek documents that setting \cw{temperature} ``will not trigger an error but will also have no effect'' & Two identical calls with \cw{temperature = 0} on a reasoning prompt & Visible derivations differed token by token, and the reasoning-token counts differed by roughly 13\%. \\
Gemini 3.1 Pro\tnote{a} & Google requires explicit \cw{tools.google\_search} to enable grounding & One call with no \cw{tools} field asking for a breaking news story from the last twenty four hours & Refusal phrase emitted. 188 reasoning tokens billed on a one sentence question (consistent with the model's Type 1 mandatory-reasoning classification). \\
Claude Opus 4.7\tnote{a} & Anthropic extended thinking is off unless the \cw{thinking} field is set & One call with no \cw{thinking} field, prompt ``Reply with exactly the number: 2+2'' & \cw{reasoning\_tokens = 0} in the returned \cw{usage} object. \\
\bottomrule
\end{tabular}
\begin{tablenotes}[flushleft]
\footnotesize
\item[(a)] Full OpenRouter model slugs. Grok 4.20 = \texttt{x-ai/grok-4.20}, GPT-5.5 = \texttt{openai/gpt-5.5}, DeepSeek R1 = \texttt{deepseek/deepseek-r1}, Gemini 3.1 Pro = \texttt{google/gemini-3.1-pro-preview}, Claude Opus 4.7 = \texttt{anthropic/claude-opus-4.7}.
\item All eight calls used the same authentication, the same gateway endpoint, and the same payload shape (modulo the controlled parameter under test) as the en-masse data collection. The verification scripts are at \texttt{\_tests/grok\_search\_default.py} (Grok) and \texttt{\_tests/verify\_other\_models.py} (the other four tests). Aggregate cost \$0.0086 USD reported by the gateway's \cw{usage.cost} field.
\end{tablenotes}
\end{threeparttable}
\end{table}

\subsection{API versus Local Execution and the Model-as-Deployed Framing}

The empirical study calls every model through the OpenRouter gateway. A reader could ask whether the same measurements would be obtained from a local installation of an open-weight model running on a researcher's hardware, and whether the gateway path is an avoidable confound on what the study claims to measure. I address the comparison directly.

The object the study measures is the model as it is deployed on the gateway through which an applied user will reach it. For the closed-weight families (Anthropic, OpenAI proper, Google Gemini, xAI Grok) there is no alternative. The weights are not public and a local execution path does not exist. The only reproducible call is the gateway call. For the open-weight families (DeepSeek, Qwen, Gemma, gpt-oss, GLM-5, NVIDIA Nemotron, Mistral) a local execution path does exist in principle. The weights are public and inference can run on the researcher's hardware. I chose the gateway path for these families as well, for two reasons. First, the substantive question of the paper concerns the model that applied users actually encounter, which is the gateway-served endpoint and not the bare weights. Second, mixing local execution for some contestants with gateway execution for the rest would introduce a measurement asymmetry across the contestant pool that is itself a confound. The closed-weight cells would carry provider-side defaults that the open-weight cells would not, and the comparison between the two halves of the pool would be contaminated by that fact.

A literature on reproducibility in LLM evaluation cautions that closed-source models are a moving target. \citet{laskar2024survey} review the practice of using such models as evaluators and note that frequent provider-side updates affect repeatability. The same caution applies to using closed-source models as contestants. I do not claim that the measurement is invariant under future provider updates. The \cw{model\_slug} resolves at call time to whatever build the provider is serving, and the \cw{native\_model} field returned by the gateway records the resolved identifier. The audit log preserves this identifier together with the \cw{gen\_id} of each call, so the exact endpoint state at the time of each measurement is recoverable.

A second literature on benchmark contamination cautions that the questions on which a model is measured may have been seen during pre-training. \citet{balloccu2024leak} document the scale of indirect data contamination in closed-source models through a systematic review. For my study the response has two parts. First, the red-herring manipulation is novel by construction. The paired treatment and control versions of every question were built for this study and do not exist in any public corpus, so any contamination of a closed-weight model with my questions is bounded to the underlying qualifying-exam item and not to the version-specific intervention that identifies the red-herring effect. Second, the identification of the red-herring effect rests on the within-item, within-model contrast across versions of the same question. The model fixed effect absorbs whatever the model knows about the item from training, and the version contrast subtracts off the level. Contamination of the underlying item would inflate the absolute level of accuracy in both arms but would not bias the estimated red-herring effect under the design.

A third concern, specific to the open-weight contestants, is precision. Local execution at the model card's stated precision (typically BF16 or FP16) is the reference. Gateway providers may serve the same weights quantized. One gateway provider states explicitly that quantization changes outputs, with the largest effects on long chains of thought. The \cw{native\_model} column and the per-call cost in the audit log let a reader audit which provider served each open-weight cell. The routing rule already excludes the providers I identified as low-output-cap. For what remains, the model-as-deployed framing absorbs the quantization choice as part of what is being measured.

The framing has a concrete deliverable. The combination of \cw{model\_slug}, \cw{native\_model}, \cw{gen\_id}, and the per-call entry in the audit log records, for every row in the analytic panel, the exact model build, the exact provider that served the request, and the gateway-internal identifier of the response. A future researcher with access to the same gateway can attempt to reproduce any individual cell. The reproducibility of an individual cell remains subject to the provider's update cadence. The reproducibility of the study as a whole rests on the panel as a finished artifact and on the audit log as the record of how it was assembled.

\subsection{Grading}
Correctness is assigned by a judge model, not by the author. The primary judge is xAI Grok 4.3 and the second, independent judge is Google Gemini 3.1 Pro. Both reason by default, and neither is in the contestant pool, so no model grades itself. The judge runs in a fresh conversation after the contestant has finished, with no memory of the model under test. The judge is given the verified answer and the reference worked solution in its own prompt, and the contestant is given neither.

On an explanation-requested version the judge returns four marks together with the model's own final answer. The four marks are whether the final answer is correct, whether the response contains an explanation, whether the reasoning shown follows a path equivalent to the reference solution, and whether the final answer is supported by the reasoning shown. I call the last of these coherence in the paper, and I reserve the word consistency for the separate cross-wave measure. On a no-explanation version the original rubric graded only whether the answer is correct, and I later regraded those responses for whether an explanation was given.

The second judge is present to test whether the finding depends on who grades. For about twelve percent of the judged cells it regrades the same response independently. Table~\ref{tab:supp_interjudge} reports the agreement between the two judges on the explanation-condition regrade subsample. They agree on almost every response, and the red-herring effect is the same size under either judge, so the result does not rest on the choice of grader. Gemini 3.1 Pro also served as a fallback on a small number of responses that the primary judge could not parse, and the audit log records which judge graded which cell.
\begin{center}\captionof{table}{Inter-judge agreement on the second-judge regrade subsample.}\label{tab:supp_interjudge}\vspace{1.5\baselineskip}
\begin{threeparttable}
\begin{tabular}{@{}l cc c@{}}\toprule
 & \multicolumn{2}{c}{Second judge, Gemini 3.1 Pro} & \\
\cmidrule(lr){2-3}
Primary judge, Grok 4.3 & Correct & Incorrect & Total \\\midrule
Correct   & 1{,}485 & 6 & 1{,}491 \\
Incorrect & 56 & 1{,}087 & 1{,}143 \\\midrule
Total     & 1{,}541 & 1{,}093 & 2{,}634 \\\bottomrule
\end{tabular}
\begin{tablenotes}[flushleft]\scriptsize
\item Each cell counts the responses on which the primary judge and the second judge return the marks in its row and column, over the 2{,}634 responses of the 2{,}750-response explanation-condition regrade subsample where both judges returned an answer-correctness verdict. The two judges agree on 97.6\% of these responses, and Cohen's $\kappa$ is 0.95.
\item Why two judges. The second judge tests whether the finding depends on who grades, so the relevant comparison is between the two judges on this subsample rather than against the headline. Estimated on this subsample, which holds only the explanation-condition responses that were regraded, the red-herring effect on answer correctness is $-0.150$ (se $0.031$) under Grok 4.3 and $-0.152$ (se $0.032$) under Gemini 3.1 Pro. The two match, so the grader does not move the result. Both are larger than the full-panel headline of $-0.123$ because this subsample is the explanation condition alone, where the main-text mechanism estimate on accuracy is already $-0.130$, and it is a small regraded fraction of the panel. Both use a linear probability model with task and model fixed effects and standard errors clustered by task.
\end{tablenotes}
\end{threeparttable}
\end{center}

\subsection{Second-Pass Grading of Spontaneous Explanation on Noexp Cells}\label{app:noexpregrade}

The grading rubric's noexp branch grades only one check, whether the answer is correct. The other three checks the exp branch applies (explanation given, reasoning correct, answer follows from the reasoning) were not run on noexp cells. The design was motivated by simplicity, since noexp prompts do not ask the model to explain. But a model can explain even when not asked. Whether it does, and whether the unprompted reasoning is correct, are economically meaningful behaviors that the original protocol discarded.

I add a second-pass grading of these three checks on every noexp cell with a non-empty response. The judge is the same Grok 4.3 used in the en-masse run, called at temperature zero through the same OpenRouter endpoint \citep{openrouter_apiref_2026}. The rubric grades three boolean fields with an NA-encoding convention.

The convention separates two semantically distinct cases. A noexp response that is just the bare answer (a phrase like ``increase'' or a single number) is fully compliant with the prompt and the three new fields encode this as \cw{explanation\_given = False}, with \cw{reasoning\_correct} and \cw{consistent} returned as JSON \cw{null} and stored as blank in the CSV. A noexp response that spontaneously contains reasoning is encoded as \cw{explanation\_given = True} with \cw{reasoning\_correct} and \cw{consistent} taking boolean values that grade the spontaneous reasoning. The convention prevents downstream analyses from treating ``the model did not reason here'' as ``the reasoning was wrong'' (two distinct quantities), and lets the analyst condition on the explanation-given indicator without ambiguity.

Two of 22,800 noexp cells could not be graded by Grok 4.3 across multiple retries. Both contained very long responses (33,000 and 36,000 characters of LaTeX-formatted derivation) from \cw{qwen/qwen3-next-80b-a3b-instruct}. I graded them with my second judge, \cw{google/gemini-3.1-pro-preview}, using the identical rubric. The audit log records the model used per cell.

Table~\ref{tab:noexpregrade} reports the counts. The substantive measurement is that 13,095 of 22,477 gradable noexp cells (58 percent) contained spontaneous reasoning despite the prompt asking only for an answer. Of those, the judge marked 56 percent as having correct reasoning and 93 percent as internally coherent (the answer follows from whatever reasoning was given, regardless of whether that reasoning is itself correct). The 323 ungradable noexp cells were truncated, errored, or returned an empty response. They are blank-by-design across the three new fields and remain missing observations.

\begin{table}[t]
\centering
\caption{Second-pass grading of the 22,800 noexp cells. The judge is the same Grok 4.3 used in the main run, at temperature zero, with an NA-encoded four-criteria rubric. Two refractory cells with very long responses were graded by Gemini 3.1 Pro as a second-judge fallback. Of the 22,477 gradable noexp cells, 58 percent contained spontaneous explanation. Among those, 56 percent had correct reasoning and 93 percent had a coherent answer-reasoning link.}
\label{tab:noexpregrade}

\begin{threeparttable}
\resizebox{\linewidth}{!}{%
\begin{tabular}{@{}lrl@{}}
\toprule
state & count & share \\
\midrule
graded with spontaneous explanation & 13{,}095 & 57.4\% of noexp \\
\quad of those, reasoning correct & 7{,}395 & 56.5\% of with-explanation \\
\quad of those, coherent answer-reasoning link & 12{,}131 & 92.6\% of with-explanation \\
graded as bare answer & 9{,}382 & 41.2\% of noexp \\
ungradable by design (truncated, errored, empty) & 323 & 1.4\% of noexp \\
\midrule
total noexp & 22{,}800 & 100.0\% \\
\bottomrule
\end{tabular}}
\begin{tablenotes}[flushleft]
\footnotesize
\item Total cost of the second-pass grading was \$50.40 USD. 22,475 cells graded by Grok 4.3 at \$47.36. Two refractory cells graded by Gemini 3.1 Pro at \$0.04 (a long-response failure mode for Grok at this endpoint). The residual is API spend during a one-time filter-bug recovery (full audit preserved). The audit log of every judge call is at \texttt{RCT\_analysis/logs/regrade\_noexp\_calls.jsonl}.
\end{tablenotes}
\end{threeparttable}
\end{table}

The augmented data set replaces \cw{RCT\_analysis/rct\_results.csv} as the canonical source for the analysis pipeline. The pre-regrade version is preserved in the audit folder. The three build scripts that produce the master analysis dataset (\cw{rct\_master.csv}) were re-run with this source, and \cw{build\_master\_dataset.py} was patched at the column-reconstruction block to propagate the noexp grading into the reconstructed numeric columns it exposes (\cw{explanation\_given\_r}, \cw{reasoning\_correct\_r}, \cw{coherent}). All pre-change versions are archived for audit. Self-tests in the build scripts and a row-by-row diff confirmed the changes are confined to the three target columns and only on noexp rows.

\subsection{Per-Model Reasoning Taxonomy}
Table~\ref{tab:supp_taxonomy} gives the reasoning type and the weight availability of all forty-two catalogued models. The reasoning type is verified per model against provider documentation, OpenRouter routing metadata, and an empirical probe. The routing metadata is a set of fields the gateway publishes on each model page, chiefly whether the model reasons when no reasoning setting is sent and whether reasoning can be disabled, which I read for all forty-two models on the same date. The four models that neither the documentation nor the metadata could settle were resolved by a probe that runs each through the study pipeline several times and classifies it by whether it emits a separate stream of reasoning tokens under a default call. The weight availability follows the NTIA availability test, verified per model card.
\begin{center}\captionof{table}{Per-model reasoning-behavior taxonomy and weight availability for all forty-two catalogued models. Type~1 emits a separate reasoning-token stream by default. Type~2 accepts a reasoning parameter but emits none under these calls. Type~3 has no reasoning mode. Weight availability, open or closed, is coded on the NTIA availability test and verified per model card. Two non-contestants serve as graders (judge) and two are held out on cost.}\label{tab:supp_taxonomy}\vspace{1.5\baselineskip}\scriptsize
\resizebox{\linewidth}{!}{%
\begin{tabular}{@{}lcc@{\hspace{1.5em}}lcc@{}}\toprule
Model & Type & Weight & Model & Type & Weight \\\midrule
DeepSeek R1 & 1 & open & Claude Opus 4.1 (held out) & 2 & closed \\
DeepSeek V4 Flash & 1 & open & Claude Opus 4.5 & 2 & closed \\
DeepSeek V4 Pro & 1 & open & Claude Opus 4.6 & 2 & closed \\
GLM-5 & 1 & open & Claude Opus 4.7 & 2 & closed \\
GPT-5.5 & 1 & closed & Claude Sonnet 4 & 2 & closed \\
GPT-5.5 Pro (held out) & 1 & closed & Claude Sonnet 4.5 & 2 & closed \\
Gemini 2.5 Flash & 1 & closed & Claude Sonnet 4.6 & 2 & closed \\
Gemini 2.5 Pro & 1 & closed & DeepSeek V3.1 & 2 & open \\
Gemini 3 Flash Preview & 1 & closed & DeepSeek V3.2 & 2 & open \\
Gemini 3.1 Flash Lite Preview & 1 & closed & GPT-5.4 & 2 & closed \\
Gemini 3.1 Pro Preview (judge) & 1 & closed & GPT-5.4 mini & 2 & closed \\
Gemini 3.5 Flash & 1 & closed & GPT-5.4 nano & 2 & closed \\
Grok 4.20 Multi-Agent & 1 & --- & Gemma 4 26B A4B & 2 & open \\
Grok 4.3 (judge) & 1 & closed & Gemma 4 31B & 2 & open \\
NVIDIA Nemotron 3 Super & 1 & open & Grok 4.20 & 2 & closed \\
Qwen3 Next 80B Thinking & 1 & open & Mistral Medium 3.5 & 2 & open \\
Qwen3.6 Plus & 1 & closed & Gemini 2.0 Flash & 3 & closed \\
Qwen3.7 Max & 1 & closed & Gemini 2.0 Flash Lite & 3 & closed \\
gpt-oss-120b & 1 & open & Qwen3 235B A22B Instruct & 3 & open \\
gpt-oss-20b & 1 & open & Qwen3 Coder 480B A35B Instruct & 3 & open \\
Claude Haiku 4.5 & 2 & closed & Qwen3 Next 80B Instruct & 3 & open \\\bottomrule\end{tabular}}
\end{center}

\subsection{Per-Model Damage and Baseline Accuracy}
A natural question is whether the more capable models are protected. Figure~\ref{fig:damage} plots each model's red-herring damage against its accuracy without the red herring, the clean measure of ability, and the answer is no. The fit is flat, so the strongest models are harmed about as much as the weakest.
\begin{figure}[htbp]
\centering
\includegraphics[width=0.8\linewidth]{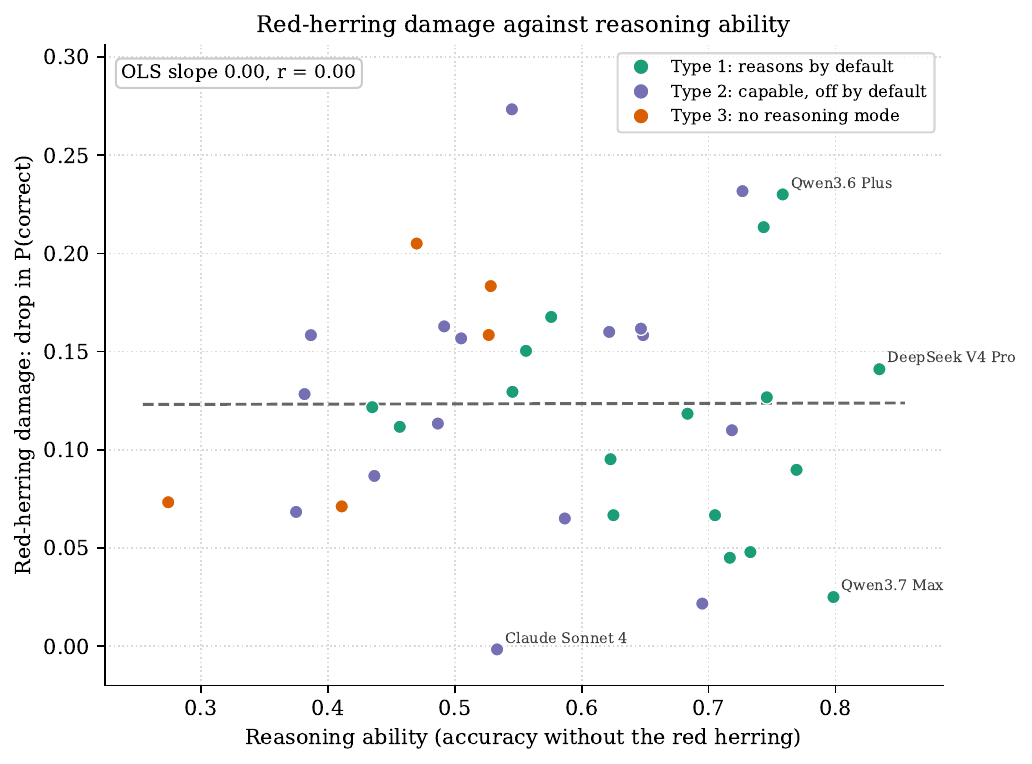}
\caption{Red-herring damage against pure ability. For each model the horizontal axis is its accuracy without the red herring, the clean measure of ability, and the vertical axis is the damage, the drop in the probability of a correct answer when the red herring is added. Points are colored by reasoning type and the dashed line is the ordinary least squares fit. The fit is flat, with a slope near zero and a correlation of $0.00$, and it sits at the pooled damage of $0.12$. Reasoning ability does not predict damage. The most accurate models are harmed as much as the least accurate, and among the strongest models the damage runs from two to twenty-three points, so robustness is idiosyncratic rather than a return to ability. Because damage is the difference between the control and treated accuracies it shares the control term with the horizontal axis, which would bias a naive fit toward a positive slope, and none appears.}
\label{fig:damage}
\end{figure}

\end{document}